\documentclass{aa} 
\usepackage{graphicx}
\usepackage[varg]{txfonts}
\usepackage{natbib}
\usepackage[version-1-compatibility]{siunitx}
\DeclareSIUnit{\Msol}{\, \mathrm{M_\odot}}
\DeclareSIUnit{\Lsol}{\, \mathrm{L_\odot}}
\DeclareSIUnit{\Msolyr}{\, \mathrm{M_\odot \, yr^{-1}}}
\DeclareSIUnit{\ar}{\mathrm{a_R}}
\newcommand{\norm}[1]{\left\lVert#1\right\rVert} 
\newcommand{\ramses}{\textsc{Ramses}}

\newcommand{\be}{\begin{equation}}
\newcommand{\ee}{\end{equation}}
\newcommand{\beal}{\begin{aligned}}
\newcommand{\eeal}{\end{aligned}}
\newcommand{\mc}{M_\mathrm{c}}

\newcommand{\rmc}{\mathrm{c}}

\def\uvec{\ensuremath{\mathbf{u}}}
\def\pveci{\ensuremath{\mathbf{p}_i}}

\usepackage{hyperref}
\bibpunct{(}{)}{;}{a}{}{,} 
\usepackage{amsmath}
\usepackage{esdiff}
\usepackage{xcolor}
\usepackage[english]{babel}
\usepackage{multirow}

\begin{document}

   \title{Collapse of turbulent massive cores with ambipolar diffusion and hybrid radiative transfer \\
   II. Outflows}

   \author{R. Mignon-Risse
          \inst{1,2}
          \and
          M. Gonz\'alez \inst{1}
          \and
          B. Commer\c con \inst{3}
          }

   \institute{AIM, CEA, CNRS, Universit\'e Paris-Saclay, Universit\'e de Paris, F-91191 Gif-sur-Yvette, France \\
         \email{raphael.mignon-risse@apc.in2p3.fr}
         \and
         Universit\'e de Paris, CNRS, Astroparticule et Cosmologie,  F-75013 Paris, France
         \and
             Centre de Recherche Astrophysique de Lyon UMR5574, ENS de Lyon, Univ. Lyon1, CNRS, Universit\'e de Lyon, 69007 Lyon, France
             }

   \date{Received ?; ?}

 
  \abstract
   {Most massive protostars exhibit bipolar outflows. Nonetheless, there is no consensus regarding the mechanism at the origin of these outflows, nor on the cause of the less-frequently observed monopolar outflows.
    }
   {We aim to identify the origin of early massive protostellar outflows, focusing on the combined effects of radiative transfer and magnetic fields in a turbulent medium.
 }
   {We use four state-of-the-art radiation-magnetohydrodynamical simulations following the collapse of massive $100\Msol$ pre-stellar cores with the  \ramses{} code. Turbulence is taken into account via initial velocity dispersion. We use a hybrid radiative transfer method and include ambipolar diffusion.
   }
   {Turbulence delays the launching of outflows, which appear to be mainly driven by magnetohydrodynamical processes. 
   We study both the magnetic tower flow and the magneto-centrifugal acceleration as possible origins. Both contribute to the acceleration and the former operates on larger volumes than the latter.
   Our finest resolution, $5$~AU, does not allow us to get converged results on magneto-centrifugally accelerated outflows.
   Radiative acceleration takes place as well, dominates in the star vicinity, enlarges the outflow extent, and has no negative impact on the launching of magnetic outflows (up to $M\,{\sim}17\, \Msol$, $L\,{\sim} \, 10^5\Lsol$). 
   We observe mass outflow rates of $10^{-5} - 10^{-4} \Msolyr$ and momentum rates of the order ${\sim}10^{-4}\Msol \, \mathrm{km \, s^{-1} \, yr^{-1}}$.
   The associated opening angles ($20-30\deg$ when magnetic fields dominate) are in a range between observed values for wide-angle outflows and collimated outflows.
      If confirmed with a finer numerical resolution at the outflow interface, this suggests additional (de-)collimating effects.
   Outflows are launched nearly perpendicular to the disk and are misaligned with the initial core-scale magnetic fields, in agreement with several observational studies.
   In the most turbulent run, the outflow is monopolar.
   }
   {Magnetic processes dominate the acceleration of massive protostellar outflows up to ${\sim}17\Msol$, against radiative processes. Turbulence perturbs the outflow launching and is a possible explanation for monopolar outflows.}


   \maketitle
%
\section{Introduction}

Massive stars form in dense environments, and one of their birth signs is the presence of (often bipolar) outflows.
Nevertheless, their large luminosities, together with the presence of magnetic fields in their birth place, has complicated the task of understanding the origin of these outflows.
Indeed, while it is now quite well accepted that low-mass protostars power magnetically-driven outflows (see e.g. \citealt{pudritz_role_2019} and references therein), the strong radiative force from massive protostars is also capable of launching outflows (\citealt{krumholz_dynamics_2009}, \citealt{kuiper_three-dimensional_2011}, \citealt{rosen_unstable_2016}, \citealt{mignon-risse_new_2020}).
Moreover, magnetic protostellar outflows rely on disk-mediated accretion, while the accretion mechanism of massive protostars has long been under debate, with additional modes such as accretion via stellar collisions \citep{bonnell_formation_1998}, radiative Rayleigh-Taylor instabilities (\citealt{krumholz_dynamics_2009}, \citealt{rosen_unstable_2016}) or dense filaments \citep{rosen_unstable_2016}.
Distinguishing between a magnetically-driven and a radiatively-driven outflow in a self-consistent way requires, at least, to solve the magnetohydrodynamics (MHD) equations coupled to radiative transfer equations.
This is the purpose of this paper.

There are now numerous observational clues on the outflow mechanism around massive protostars.
Several works agree on a clear correlation between the source radio luminosity up to $10^5 \Lsol$, the core mass and the outflow momentum rate (\citealt{anglada_radio_1992}, \citealt{cabrit_co_1992}, \citealt{beuther_massive_2002}, see the review by \citealt{anglada_radio_2018}).
This correlation is found to continue to high luminosities, i.e. high-mass protostars (core masses ranging from hundreds to thousands of solar masses, \citealt{beuther_massive_2002}).
Collimated jets, a common feature of low-mass star formation are also observed around massive young-stellar objects (e.g., \citealt{moscadelli_water_2005}).
As there is no clear correlation between the lines luminosity in the observed winds and the stellar photospheric luminosity \citep{cabrit_forbidden-line_1990}, the outflow mechanism likely originates from the disk and not from the star itself.
It is also an additional argument in favor of disk accretion.
Outflows seem to have an onion-shell structure \citep{cabrit_co_1990}, with large velocities close to the outflow axis and a decreasing velocity as gas is located further away from the axis, in agreement with the MHD disk wind theory (\citealp{blandford_hydromagnetic_1982}, \citealt{spruit_magnetohydrodynamic_1996}).
The cavity walls formed by the outflows have been revealed, e.g. with the Subillimeter Array (SMA) in the GGD27 complex which hosts a ${\sim}4 \Msol$ protostar powering a thermal radio jet (\citealt{fernandez-lopez_millimeter_2011}, \citealt{girart_circumestellar_2017}).
Evidence of precession is associated to this source's molecular outflows \citep{fernandez-lopez_multiple_2013}, similarly to those around low-mass protostars (e.g., \citealt{de_valon_alma_2020}).
\cite{hirota_disk-driven_2017} have found signs of rotation within an outflow as well as the presence of a disk with the Atacama Large Millimeter/submillimeter Array (ALMA), around a ${\sim}15\Msol$ source \citep{ginsburg_keplerian_2018}.
These bring evidence of outflows originating from a MHD disk wind for high-mass protostars, similarly to their low-mass counterparts.
Finally, the outflow orientation with respect to the core-scale magnetic fields could give us insights on the role of magnetic fields. \cite{zhang_alma_2016} find its main axis direction does not seem correlated to the magnetic field orientation. It could indicate that the disk orientation may not be governed by magnetic braking but by other dynamical interactions, as in multiple systems. However, the magnetic braking efficiency depends on the orientation between magnetic fields and angular momentum (\citealt{hennebelle_disk_2009}, \citealt{joos_protostellar_2012}), hence this needs to be further investigated.
Because magnetic braking would reduce the disk size, constraints on disk geometry can help us to identify the exact role of magnetic fields in massive star formation.

Indeed, disk accretion is the most favored accretion mechanism for stars of all masses, and their presence around massive protostars has growing evidences (see \citealt{beltran_disks_2020} for an up-to-date review).
Early theoretical works have shown that accretion disks can power fast ($\gtrsim \, 100 \mathrm{ \,km \, s^{-1}}$) jets by magneto-centrifugal acceleration (\citealp{blandford_hydromagnetic_1982}, \citealt{pudritz_centrifugally_1983}, \citealt{pelletier_hydromagnetic_1992}) or 
slow (${\sim}\, 1-10 \mathrm{ \, km \, s^{-1}}$) magnetic-pressure-gradient driven tower flows (\citealp{lynden-bell_magnetic_1996}, \citealt{lynden-bell_why_2003})
by twisting the field lines and accumulating enough toroidal magnetic field.
The former is characterized by a very collimated structure and a magnetic field whose poloidal component is dominant at the launching region (the inner disk regions).
Gas is accelerated along the field lines (no magnetic force) by the centrifugal acceleration until its motion becomes super-Alfv\'enic so that field lines lag behind its conserved rotation motion and self-collimates by the magnetic tension force.
The magnetic tower flow gives rise to a wide-angle outflow and is dominated by the toroidal component in the launching region (as fields lines are wound-up by the disk) and in the entire flow.
Gas is accelerated perpendicular to the twisted field lines by the Lorentz acceleration.
For a review on the numerical advances regarding these processes we refer the reader to \cite{pudritz_disk_2007}, and for their role in star formation to \cite{pudritz_role_2019}.

The presence of these two types of magnetic outflows, namely magneto-centrifugal and magnetic tower flows, has been confirmed in numerical simulations.
Both outflows have been obtained under the ideal MHD approximation, in the low-mass regime (\citealt{hennebelle_magnetic_2008} and \citealt{banerjee_outflows_2006}), later-on in the high-mass regime (\citealt{hennebelle_collapse_2011},\citealt{seifried_magnetic_2012}).
Using sub-AU resolution 3D calculations of massive core collapse, \cite{banerjee_massive_2007} obtained the early bipolar outflows but do not follow the calculation after a star has formed.
Relaxing the ideal MHD approximation, the question has been tackled with the inclusion of Ohmic dissipation by \cite{matsushita_massive_2017} and \cite{kolligan_jets_2018}.
\cite{matsushita_massive_2017} used 3D nested grids with equatorial symmetry to reach very high-resolution ($0.8$~AU).
They find that the ratio between the mass outflow rate and the mass accretion rate is nearly constant  throughout the stellar mass spectrum, indicating a common launching mechanism, in line with the observational constraints (see e.g., \citealp{wu_study_2004}).
Including ambipolar diffusion, \cite{commercon_discs_2021} (hereafter, C21) obtained qualitatively similar results.
With a 2D spherical grid, \cite{kolligan_jets_2018} studied the launching of both types of outflows with an even higher resolution ($0.09$~AU) and Ohmic dissipation around a massive protostar.
They found that only a spatial resolution of $\lesssim 0.17$~AU at $1$~AU could provide numerically-converged results on the magneto-centrifugal jets, while distinguishing both types of outflows was very difficult in their low-resolution run.
The conclusions from these works are twofold.
First, the outflow mechanisms during low- and high-mass star formation could be the same.
Second, sub-AU resolution is required to obtain converged results on the magneto-centrifugal jets.
Nonetheless, these MHD-oriented works have neglected a key ingredient at play in massive star formation: radiative transfer.

Many numerical studies have shown the production of radiative outflows in a radiation-hydrodynamical framework, using the popular flux-limited diffusion (FLD) method \citep{levermore_flux-limited_1981}.
However, stellar radiation propagates along rays, hence it requires a method capable of conserving its directionality.
Moreover, the dust opacities are very sensitive to the radiation frequency, and stellar radiation is ultraviolet-like radiation while dust emission is infrared.
The desired numerical method should track this frequency information, from stellar radiation emission to absorption by the surrounding dust.
Otherwise, the opacity of the first absorption event of stellar radiation is underestimated and the radiative force along with it \citep{owen_radiative_2014}.
Numerous irradiation implementations have been designed for massive star formation (\citealt{kuiper_fast_2010},\citealt{rosen_hybrid_2017}, \citealt{mignon-risse_new_2020}) or for the physical structure of protoplanetary disks (\citealt{flock_radiation_2013}, \citealt{ramsey_radiation_2015}, \citealt{gressel_global_2020},
\citealt{melon_fuksman_two-moment_2021}).
Radiative cavities have been found to form after the central star has reached ${\sim}10\Msol$, so the corresponding luminosity can drive a radiative force capable to overcome the gravitational force (and ram pressure).
Radiative outflows are characterized by velocities of ${\sim}10-20 \mathrm{\, km \, s^{-1}}$ (\citealt{rosen_unstable_2016}, \citealt{mignon-risse_new_2020}).
Nonetheless, the stellar radiative acceleration appears to be too weak to explain the momentum rate of bipolar outflows observed around protostars of all masses (\citealt{lada_cold_1985}, \citealt{cabrit_co_1992}), by $1-2$ orders of magnitude.

The first implementations of both a radiative transfer method and an MHD solver have been targeted towards the physics of fragmentation (see e.g., \citealt{commercon_collapse_2011}, \citealt{peters_interplay_2011}).
Only few works have focused on the co-launching of radiative and magnetic outflows, since it requires a hybrid radiative transfer method (not to underestimate stellar feedback), (non-ideal) MHD (to obtain a realistic disk and self-consistent outflows) and sub-AU resolution.
To circumvent this difficulty, subgrid models have been used to mimic protostellar outflows and found to dominate over the radiative ones \citep{rosen_role_2020} and to enhance the flashlight effect \citep{kuiper_protostellar_2015}.

Two dedicated works have investigated the impact of stellar radiation on the launching and structure of magnetic outflows.
On the one hand, including photoionizing radiation (but no radiative force) and in the ideal MHD frame, \cite{peters_interplay_2011} have shown that the development of $\mathrm{H_{II}}$ regions perturbs the magnetic fields topology and weakens the tower flow (the typical launching radius for magneto-centrifugal outflows was not resolved though).
Nonetheless, \cite{peters_morphologies_2014} show that the CO emission associated to ionization feedback could not reproduce observations.
On the other hand, \cite{vaidya_jet_2011} have focused on the collimation of magnetic jets in axisymmetric setups with ideal MHD and prescriptions for radiative forces. They observe that line-driven radiation force from a $30\Msol$ star starts to compete with magnetic forces for disk field strengths $\lesssim 5$~G at $r=1$~AU and moderately reduces the jet collimation but do not disrupt the magnetic field geometry.
Including ambipolar diffusion, using the FLD method and an aligned rotator in their initial setup of a collapsing massive core, C21 have found the outflows to be launched magnetically, while the Lorentz force dominates over the radiative force by several orders of magnitude.
C21 have shown that early massive protostellar outflows are magnetic, but their treatment of radiative transfer underestimates the radiative force. Hence, we explore whether this conclusion remains valid when including a more realistic model for irradiation, and turbulence.

In addition, it has been shown that the massive star radiative force could create cavities.
The question whether it would dominate over magnetic forces at launching outflows, or if it would be sufficient to disturb the field geometry, preventing the launching of MHD outflows, has to be assessed in a self-consistent framework.
In this work, we use the numerical simulations presented in \cite{mignon-risse_collapse_2021} (hereafter Paper I), which include both a hybrid radiative transfer method and non-ideal MHD effects (ambipolar diffusion here). 
We extend the work of C21 which has been performed with the FLD method in a non-turbulent medium, focusing on the magnetic effects.
In the present study, four runs are considered with various levels of turbulence and magnetic fields, aiming at identifying the outflow origin and its dependency on environmental conditions.
We will finally investigate to what extent our results compare with current observational constraints on massive protostellar outflows and on the disk-outflow and outflow-magnetic field alignments.

This paper is organized as follows: numerical methods are summarized in Sect.~\ref{sec:model} (we refer the reader to Paper I for more details), Sect.~\ref{sec:outflows} is dedicated to the study of the outflows, focusing on their origin, and in Sect.~\ref{sec:outflowsobs}, we compare several of their properties (e.g. opening angle, momentum rate) with observations, trying to assess how realistic our numerical results are and, consequently, if the identified mechanism is a robust candidate for massive protostellar outflows.

\section{Methods}
\label{sec:model}

\subsection{Setup}

\begin{table*}
\centering
\caption{Initial conditions of the four runs.}
\label{table:tests} 
\begin{tabular}{c | c c | c c c c  c | c c c} 
	\hline\hline
	 Model & $M_\mathrm{c}$ [$\Msol$] & \, $R_\mathrm{c}$ [pc] & $\rho(r)$ \, & $T_\mathrm{c}$ [K] & \, $E_\mathrm{th}/E_\mathrm{grav}$ [\%] & \, $M_\mathrm{plateau}/M_\mathrm{Jeans}$  & $\Omega$ [$\mathrm{s^{-1}}$] & $\mathcal{M}$ & \, $\mathcal{M}_\mathrm{A}$ & \, $\mu$  \\ \hline \hline
	\textsc{NoTurb}  & \multirow{4}{*}{$100$} & \multirow{4}{*}{$0.2$}   &  \multirow{4}{*}{$\rho_\mathrm{c}/(1+r/r_\mathrm{c})^2$}  &  \multirow{4}{*}{$20$} &  \multirow{4}{*}{$6.2$} &  \multirow{4}{*}{$13.6$}	&  \multirow{4}{*}{$9.5 \times 10^{-15}$} &  0	 & 	0	  &  5 \\  \cline{1-1}
    \textsc{SupA}	 &  					& 				  &												  &				 & 					 & 					& 							&  0.5   & 1.4		 & 5 \\ \cline{1-1}
	\textsc{SupAS}	&				  	& 				  &												  &				& 					 & 					 & 							& 	2 &	 5.7  & 5  \\  \cline{1-1}
	\textsc{SubA}    & 					 & 				   &												 &				 & 					& 					& 							&  0.5    &   0.57   & 2  \\ \hline 
\end{tabular}
{\raggedright \footnotesize{ \textbf{Notes.} $M_\mathrm{c}$, $R_\mathrm{c}$ and $T_\mathrm{c}$ are the pre-stellar core mass, radius, and temperature, respectively. $\rho(r)$ is the density as a function of the radius, $E_\mathrm{th}/E_\mathrm{grav}$ is the ratio between the thermal and the gravitational energies of the core. $M_\mathrm{plateau}$ and $M_\mathrm{Jeans}$ are the total mass in the central plateau and the local Jeans mass, respectively. Their ratio gives the number of Jeans masses contained within the central plateau, as an additional measurement of the thermal support versus gravity. $\Omega$ is the solid-body rotation frequency of the core. $\mathcal{M}$ and $\mathcal{M}_\mathrm{A}$ are the Mach number and Alfv\'enic Mach number, respectively. $\mu$ is the mass-to-flux ratio divided by the critical mass-to-flux ratio.} \par}
\end{table*}

We use the suite of four radiation-magnetohydrodynamical simulations presented in Paper I (including four lower-resolution runs).
Let us summarize their main characteristics.
These are run with the adaptive-mesh refinement code \ramses{} (\citealt{teyssier_cosmological_2002}, \citealt{fromang_high_2006}).
Non-ideal MHD is accounted for in the form of ambipolar diffusion \citep{masson_incorporating_2012} and we use the hybrid radiative transfer method  \citep{mignon-risse_new_2020}, i.e. a M1 closure relation \citep{levermore_relating_1984} to treat stellar radiation from the primary sink, while all radiation emitted otherwise is modeled with the Flux-Limited Diffusion method (FLD, \citealt{levermore_flux-limited_1981}).
An ideal equation of state is employed to relate the specific internal energy to the dust-gas mixture (with $1\%$ dust-to-gas ratio) temperature.
In this framework, we follow the collapse of a $M_\mathrm{c}=100\, \Msol$ pre-stellar core of radius $R_\mathrm{c}=0.2$~pc.
The density profile follows the relation $\rho(r) = \rho_\mathrm{c}/(1+r/r_\mathrm{c})^2$, with $\rho_\mathrm{c}{\sim}7.7\times 10^{-18} \, \mathrm{g \, cm^{-3}}$ and $r_\mathrm{c}=0.02$~pc the size of the central plateau, which contains about $15 \Msol$.
The initial, uniform temperature is $T_\mathrm{c}=20$~K, resulting in a ratio between the thermal and gravitational energies $E_\mathrm{th}/E_\mathrm{grav}=6.2\%$. The central plateau contains $M_\mathrm{plateau}/M_\mathrm{Jeans}=13.6$ Jeans masses. Solid-body rotation is imposed with a rotational frequency $\Omega \, {\approx} \,9.5\times 10^{-15}$, which gives a ratio between the associated rotational energy and gravitational energy of ${\approx}1\%$.
A velocity field consistent with a turbulent medium is initialized, whose amplitude is set by the turbulent Mach number, which varies between $0$ and $2$.
A uniform magnetic field is set aligned with the $x-$axis, with a mass-to-flux to critical-mass-to-flux ratio \citep{mouschovias_note_1976} $\mu=2$ (strong magnetic fields) or $\mu=5$ (moderate).
Our runs are labeled as follows: runs \textsc{NoTurb} (Mach number $\mathcal{M} = 0$), \textsc{SupA} ($\mathcal{M} = 0.5$), and \textsc{SupAS} ($\mathcal{M} = 2$) have $\mu=5$, and run \textsc{SubA} ($\mathcal{M} = 0.5$) has a stronger magnetic field ($\mu=2$) corresponding to sub-Alfv\'enic turbulence. Those physical parameters are given in Table~\ref{table:tests}.

Sink particles are introduced at the finest level, which corresponds to a physical resolution of $5$~AU ($10$~AU in the low-resolution runs, hereafter referred to as "LR" runs).
They accrete material in a volume of radius $20$~AU ($40$~AU for the LR runs).
They follow evolutionary tracks \citep{kuiper_simultaneous_2013} based on their mean accretion rate and mass, that give the corresponding radius, luminosity, hence effective temperature.
Accordingly, radiative energy is injected in the central oct of the sink volume, either with the M1 method (primary sink) or with the FLD method (other sinks).
Gas and radiation are decoupled within the primary sink in order to model the escape of photons with the M1 module (see the discussion in Paper I).

\subsection{Analysis: outflow properties}
\label{sec:diskdef}

Section~\ref{sec:outflows} is dedicated to the study of the outflows.
We are looking at potentially fast (${\gtrsim}\, 10 \mathrm{\, km \, s^{-1}}$) outflows but we do not want to extract very biased properties by only selecting their higher-velocity component.
Instead, we identify outflow on a cell-by-cell basis as follows.
To be considered as part of an outflow, the radial speed within a cell must exceed the escape speed: $v_r > v_\mathrm{esc}$ with $v_\mathrm{esc}= \sqrt{2 G \mathrm{M_\star}/r}$, where $r$ is the distance to the central star of mass $\mathrm{M_\star}$.
The velocity component perpendicular to the disk plane $v_\perp$ must exceed a threshold of $0.8 \mathrm{\, km \, s^{-1}}$.
This value corresponds to the maximal velocity introduced in our turbulent initial conditions in runs \textsc{SupA} and \textsc{SubA}. 
We choose this value rather than that implied by Run \textsc{SupAS} (${\approx}\,3 \mathrm{km \, s^{-1}}$) in order to minimize the bias towards high-velocity gas when computing the outflow properties and because the outflow in Run \textsc{SupAS} is weak and transient in a highly dynamical medium, making robust measurements difficult.
Taking the component perpendicular to the disk strengthens this criterion, so that potential thermal-pressure-driven, radiative-pressure-driven or interchange-instability-driven flows (see Paper I) occurring at the disk edge, parallel to the disk plane, are not counted as outflows.
Thanks to this process, we can easily obtain the mean properties of the outflow.
To go further and extract its geometry, we developed the method below.

\begin{figure*}
\centering
    \includegraphics[width=15cm]{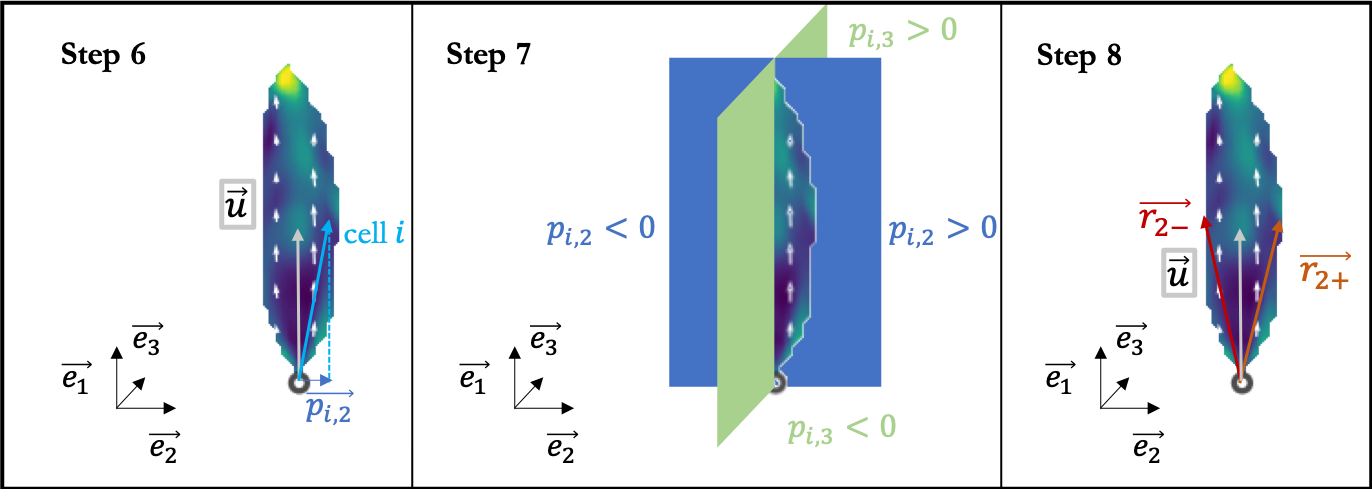}
    \caption{Outflow selection slice perpendicular to the disk in run \textsc{NoTurb}, showing three of the eight steps to compute its opening angle. It displays the projection vector $\pveci$ associated to cell $i$ (Eq.~\ref{eq:pveci}, left panel), the four subselections based on this projection (middle panel) and the geometric center vector $\uvec$ and two of the outermost positions vectors $\mathbf{r}_{2,\pm}$ used to compute the opening angle (Eq.~\ref{eq:opangle}, right panel). The circle shows the sink position, that we consider as the coordinates origin.}
    \label{fig:openingscheme}
\end{figure*}

We present here our method to extract the outflow opening angle, trying not to make strong assumptions on the outflow geometry (e.g., conical, strictly perpendicular to the disk or to the axes, axisymmetric).
Looking at bipolar outflows, we distinguish two components, each one located on one side of the disk plane, and compute their properties individually.
We consider the primary sink as the origin, call $\mathbf{r}_\mathrm{i}$ the position vector of the cell of index $i$ and create the basis $(\mathbf{e}_1,\mathbf{e}_2,\mathbf{e}_3)$ where $\mathbf{e}_1$ is colinear to the angular momentum vector $\mathbf{J}$ and $\mathbf{e}_2,\mathbf{e}_3$ are in the disk plane. For this computation, the angular momentum is taken as
$\mathbf{J} = \int_{r<10^3 \mathrm{AU}} \mathbf{r} \times \rho \mathbf{v} \, \mathrm{d}V $ where $\mathbf{r}$ and $\mathbf{v}$ are the position and velocity vectors, respectively.
\begin{enumerate}
    \item As detailed above, first select all cells with $v_r >  v_\mathrm{esc}$ and $v_\perp> 0.8 \mathrm{\, km \, s^{-1}}$.
    \item For each cell in the outflow selection, compute the dot product between the position vector $\mathbf{r}_i$ and $\mathbf{J}$ and create two sub-selections to distinguish the two cases: $\mathbf{r}_i \cdot \mathbf{J} > 0$ ("above" the disk) and $\mathbf{r}_i \cdot \mathbf{J} < 0$ ("below") that we will refer to as "A" and "B" outflows. Now we focus on one sub-selection between the two, i.e. one outflow.
    \item Define the position vector of geometric center (see Fig.~\ref{fig:openingscheme}) 
    \be
    \uvec = \frac{\sum_{i} \mathbf{r}_i \, dV_i}{\sum_i dV_i},
    \ee
    where $dV_i$ is the volume of the cell of index $i$.
    We observe transient clumps of denser gas being ejected within the outflows so taking the barycenter instead of the geometric center would lead to more variability and difficulty in interpreting the outcomes.
    \item We get the distance between the sink and the geometric center. We use this value as a sphere radius centered on the outflow geometric center and remove cells located outside the sphere : this acts as a connectivity criterion. In binary systems, we also exclude cells located within an orbital separation of the secondary star.
    \item Following the methodology of \cite{cabrit_co_1992}, we get the distance of the most distant outflow cell $R_\mathrm{outflow}$ and the volume-averaged velocity $v_\mathrm{outflow}$ of the selection to compute the outflow momentum rate
    \be
    F_\mathrm{outflow} = \frac{v_\mathrm{outflow}^2 \sum_i \rho_i \, dV_i}{R_\mathrm{outflow}}.
    \label{eq:foutflow}
    \ee
    This corresponds to the required force to accelerate the flow from a null velocity to the characteristic velocity $v_\mathrm{outflow}$ in a time scale $R_\mathrm{outflow}/v_\mathrm{outflow}$.
    \item Compute the projection $\pveci$ (left panel of Fig.~\ref{fig:openingscheme}) of the cell position vector perpendicular to the position vector of the geometrical center \uvec
    \be
    \pveci =\mathbf{r}_i - \frac{\mathbf{r}_i \cdot \uvec}{ \norm{\uvec}^2} \uvec.
    \label{eq:pveci}
    \ee
    \item Create four sub-selections $\mathbf{p}_{i,2}>0$, $\mathbf{p}_{i,2}<0$, $\mathbf{p}_{i,3}>0$ and $\mathbf{p}_{i,3}<0$ (middle panel of Fig.~\ref{fig:openingscheme}). The subscripts $2$ and $3$ denote the basis vectors $\mathbf{e}_2$ and $\mathbf{e}_3$, respectively.
    \item In the $\mathbf{p}_{i,2}>0$ sub-selection, identify the cell with $\norm{\mathbf{p}_i} = \max_i \left(\norm{\mathbf{p}_{i,2}}\right)$; its position vector is labeled $\mathbf{r}_{2,+}$. This corresponds to the outermost cell in the positive $\mathbf{e}_2$ direction. We perform the same step for $\mathbf{p}_{i,2}<0$ (outermost cell in the negative $\mathbf{e}_2$ direction), $\mathbf{p}_{i,3}>0$ and $\mathbf{p}_{i,3}<0$, and obtain $\mathbf{r}_{2,-}$, $\mathbf{r}_{3,+}$, $\mathbf{r}_{3,-}$ (right panel of Fig.~\ref{fig:openingscheme}).
    \item We define the outflow opening angle $\theta_\mathrm{outflow}$ as the average of the four angles between $\uvec$ and $\mathbf{r}_{2,+}$, $\mathbf{r}_{2,-}$, $\mathbf{r}_{3,+}$ and $\mathbf{r}_{3,-}$, respectively, i.e.
    \be
    \theta_\mathrm{outflow} = \mathrm{mean} ( \arccos \left( \frac{\mathbf{r}_{2-3,\pm} \cdot \uvec }{\norm{\mathbf{r}_{2-3,\pm}} \norm{\uvec}}  \right) \times 2 ),
    \label{eq:opangle}
    \ee
    where the factor $2$ arises because the four angles correspond to semi-opening angles.
\end{enumerate}
Let us note that, by projecting the cell positions onto the disk plane $(\mathbf{e}_2,\mathbf{e}_3)$, we implicitly assume that the outflow is perpendicular to the disk.
Since this is not generally valid, our resulting opening angle becomes less accurate as the misalignment between the outflow and $\mathbf{j}$ increases (see Sect.~\ref{sec:alignoutflow1}).

\section{Results: outflow launching mechanism and observable properties}
\label{sec:outflows}

\subsection{Analytical estimate of the origin}

We aim at studying the candidates for driving  bipolar outflows: radiative acceleration, magnetic tower flow, and magneto-centrifugal acceleration.
While modeling the latter requires strong assumptions on the magnetic field topology, we choose to compare analytically the radiative and magnetic pressure-driven accelerations.

The radiative and magnetic-pressure-gradient accelerations are respectively defined as
$a_\mathrm{rad} = \kappa F/\mathrm{c} = \kappa L / 4 \pi r^2 \mathrm{c}$ where $\kappa$ is the dust-and-gas mixture opacity, $F$ is the radiative flux coming from the star, $L$ is the stellar luminosity, $r$ is the distance to the star, and $a_\mathrm{pmag} = 1/\rho \nabla P_\mathrm{mag}= 1/\rho \nabla B^2/2$.
In the ideal MHD regime, $B\varpropto \rho^{2/3}$ and $\rho \varpropto r^{-2}$ from our initial conditions, so $B \varpropto r^{-4/3}$. 
It follows that the acceleration due to the magnetic pressure gradient can be approximated as
\be
\begin{aligned}
\frac{1}{2\rho} \diffp{B^2}{r} &= 
\frac{- 4}{3} \frac{B_\mathrm{0}^2}{r_\mathrm{0} \rho_\mathrm{0}} \left( \frac{r}{r_\mathrm{0}} \right)^{-5/3} \\
\end{aligned}
\ee
Now comparing the radiative and magnetic accelerations absolute values and deducing the luminosity for the radiative acceleration to overcome the magnetic acceleration, one obtains
\be
\begin{aligned}
a_\mathrm{rad} \equiv \frac{\kappa L}{4 \pi r^2 \rmc} &> \norm{\frac{1}{2\rho}\diffp{B^2}{r}} \equiv a_\mathrm{pmag}, \\
L&> \frac{16\pi }{3} \rmc \rho_\mathrm{0}^{-1}  r_\mathrm{0} B_\mathrm{0}^2  \kappa_\mathrm{0}^{-1} \left( \frac{r}{r_\mathrm{0}} \right)^{1/3} \left( \frac{\kappa}{  \kappa_\mathrm{0} }\right)^{-1} \\
&\gtrsim 2 \times 10^4 \Lsol \left( \frac{r}{50 \mathrm{AU}} \right)^{1/3} \left( \frac{\kappa}{ 50 \, \mathrm{cm^2 \, g^{-1}} }\right)^{-1} \\
& \quad \quad \quad \quad \quad \quad \quad \quad \left( \frac{\rho_\mathrm{0}}{10^{-15} \mathrm{g\, cm^{-3}}} \right) \left( \frac{B_\mathrm{0}}{0.1 \mathrm{G}} \right)^2,
\end{aligned}
\label{eq:arapmag}
\ee
taking $r_\mathrm{0} =50$~AU, $B_\mathrm{0}=0.1$~G, $ \kappa_\mathrm{0}=50 \, \mathrm{cm^2 \, g^{-1}}$ (the gray opacity to stellar radiation, considering an effective temperature of $4000$~K), $\rho_\mathrm{0}= 10^{-15} \mathrm{g \, cm^{-3}}$ as references, after $r_\mathrm{0}$ has been fixed.
From this equation, we can anticipate a change of regime from magnetic-dominated to radiation-dominated outflows as the protostellar luminosity increases, but only at small to intermediate scales.
Indeed, Eq.~\ref{eq:arapmag} shows that the radiative acceleration decreases more rapidly with the distance than the magnetic acceleration, so that, at large distances, magnetic tower flow is the dominant mechanism.
This analysis remains valid as long as the two components do not interact with each other.
Actually, the radiative force can push on the field lines and perturb the field topology \citep{vaidya_jet_2011}, while the tower flow dense parts can shield the rest of the outflow from stellar radiation (see the dense gas in the southern outflow, Fig.~\ref{fig:rho_outflow}).
More generally, the previous formulation is no longer valid for $r>1/(\kappa \rho_\mathrm{outflow})$ (optically-thick outflow), except to show that the radiative acceleration is overwhelmed by magnetic-pressure gradient.

\begin{figure*}
\centering
    \includegraphics[width=18cm]{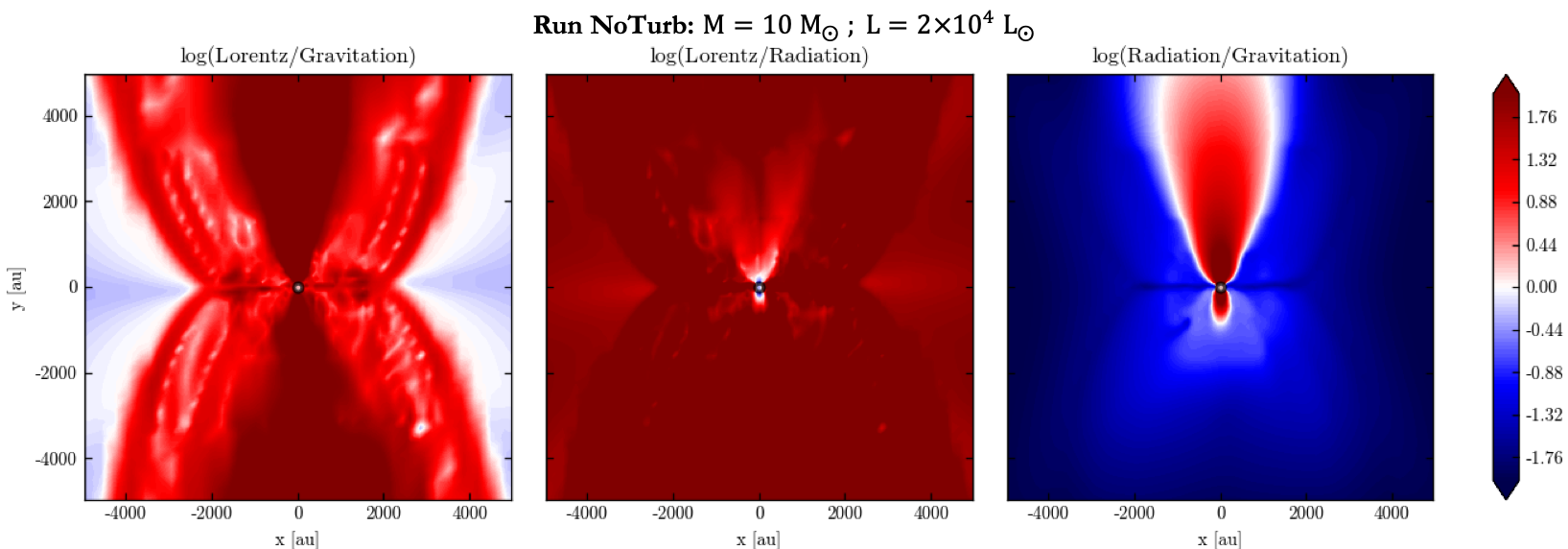}\\
    \includegraphics[width=18cm]{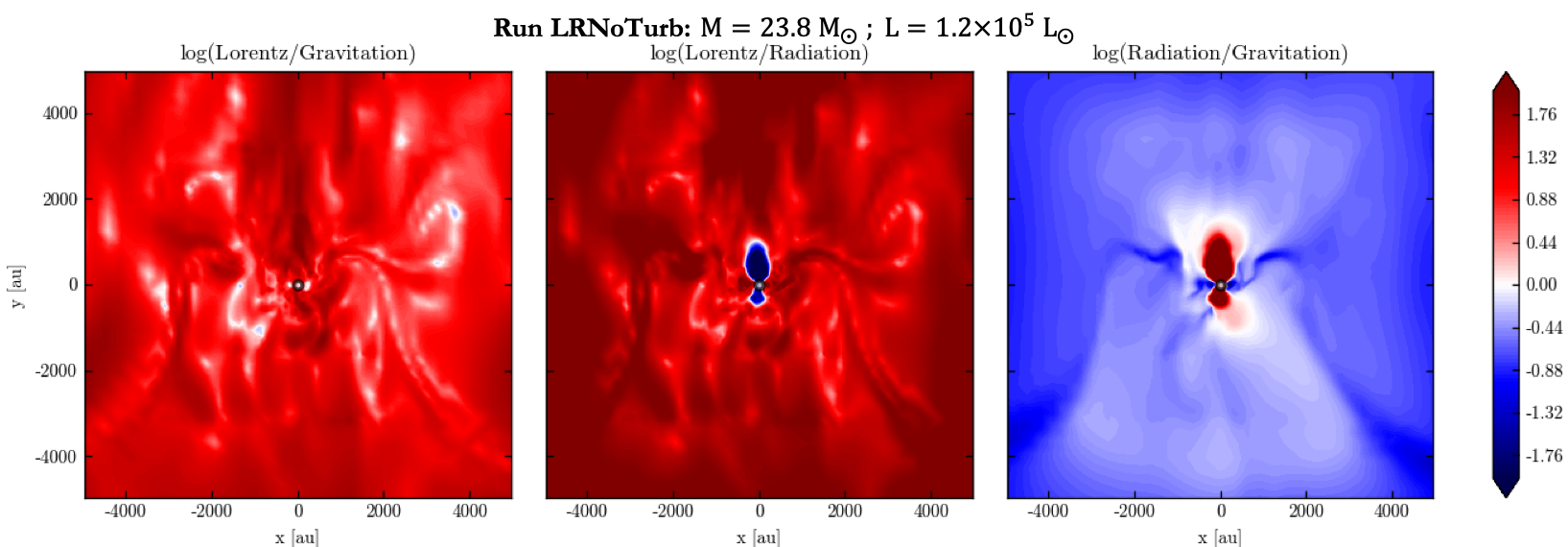}
    \caption{Slices of $10000$~AU of three forces ratios when $M=10\Msol$ ($L=2\times 10^4\Lsol$) in run \textsc{NoTurb} (top) and when $M=23.8\Msol$ ($L=1.2\times 10^5\Lsol$) in run \textsc{LRNoTurb}. Left panels: Lorentz against gravitational acceleration; middle panels: Lorentz acceleration against radiative acceleration; right panels: radiative acceleration against gravitational acceleration. Lorentz acceleration dominates over the radiative acceleration everywhere but close to the protostar.}
    \label{fig:forcesratio}
\end{figure*}

\subsection{Fiducial case: run \textsc{NoTurb}}

We start by analyzing the non-turbulent run \textsc{NoTurb} as our fiducial case. Eventually it will be compared with the study of C21 (their run \textsc{MU5AD}, with $\mu=5$).

\subsubsection{Radiative acceleration versus Lorentz acceleration}

Let us identify which of the two forces dominates when the star becomes massive ($\gtrsim 8\Msol$), in run \textsc{NoTurb} for simplicity. Figure~\ref{fig:forcesratio} shows slices perpendicular to the disk plane of the ratios $a_\mathrm{Lor}/a_\mathrm{grav}$ (left panel), which are the Lorentz and gravitational accelerations, respectively, $a_\mathrm{Lor}/a_\mathrm{rad}$ (middle panel) and $a_\mathrm{rad}/a_\mathrm{grav}$ (right panel). The snapshots are taken when the central star is $10\Msol$ and $L=2\times 10^4 \Lsol$ in run \textsc{NoTurb} and when $M=23.8\Msol$ and $L=1.2\times 10^5 \Lsol$ in run \textsc{LRNoTurb}.
The radiative acceleration $a_\mathrm{rad}$ is the total (i.e. M1 and FLD) radiative acceleration.
Run \textsc{LRNoTurb} allows us to reach a higher stellar mass and therefore a larger luminosity.
One can clearly see that both the Lorentz force and the radiative force contribute to the gas acceleration in the outflow, as they exceed the gravitational force.
Interestingly, in run \textsc{NoTurb} the radiative force contribution is very asymmetric with respect to the disk plane.
This is due to the density distribution not being symmetric, with denser gas in the southern direction stopping stellar radiation propagation, while the northern direction is particularly optically-thin at this time step.
We briefly discuss this asymmetry below.
The extent of the radiatively-dominated region is more constant with time in run \textsc{LRNoTurb}.
Indeed, it reflects a fundamental problem when modeling radiative transfer: if the photon mean free path is not resolved, absorption is overestimated.
Hence, there is more absorption in run \textsc{LRNoTurb} (a factor ${\approx}2$ at $L=2\times 10^4\Lsol$ in both runs).
We measured the absorption by taking the photon density as a function of the distance to the sink to derive an absorption factor, assuming exponential decay and after correcting for geometrical dilution. 
This difference in absorption explains why, despite a larger stellar luminosity than in the run \textsc{NoTurb} snapshot, radiation does not propagate further away.
As shown in the middle panel of Fig.~\ref{fig:forcesratio}, the Lorentz acceleration dominates the radiative acceleration everywhere but in the vicinity of the star (closer than ${\approx}300$~AU in run \textsc{NoTurb}).
In the meantime run \textsc{LRNoTurb} illustrates the stronger radiative force with a more extended zone where radiative force dominates over Lorentz force.
The center panel and right panel show very similar features, revealing that the radiative force domination is limited by absorption in run \textsc{LRNoTurb}, while it is mainly limited by geometrical dilution (inherent to an optically-thin channel) in run \textsc{NoTurb}.
To conclude, the Lorentz force dominates over the radiative force up to a stellar mass of ${\sim}20\Msol$.

\subsubsection{Radiative acceleration: FLD versus M1}

\begin{figure}
    \includegraphics[width=8cm]{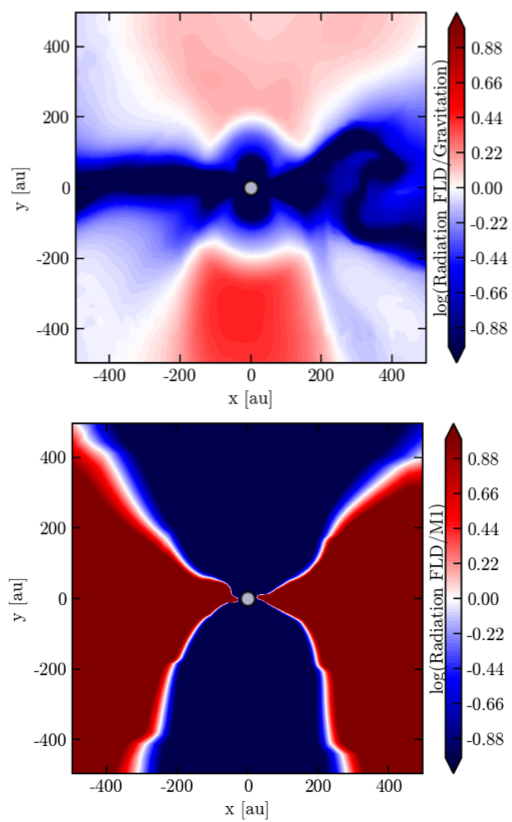}
    \caption{Slice of $1000$~AU showing the ratio of the FLD radiative acceleration to the gravitational acceleration (top) and to the M1 radiative acceleration (bottom), perpendicular to the disk plane in run \textsc{NoTurb}, when $M=12.7\Msol$. Below $M=10\Msol$ the FLD radiative acceleration rarely dominates the gravitational acceleration. The outflow region is dominated by M1 acceleration and the disk region is dominated by the FLD acceleration.}
    \label{fig:fradfldm1}
\end{figure}

\begin{figure}
    \includegraphics[width=8cm]{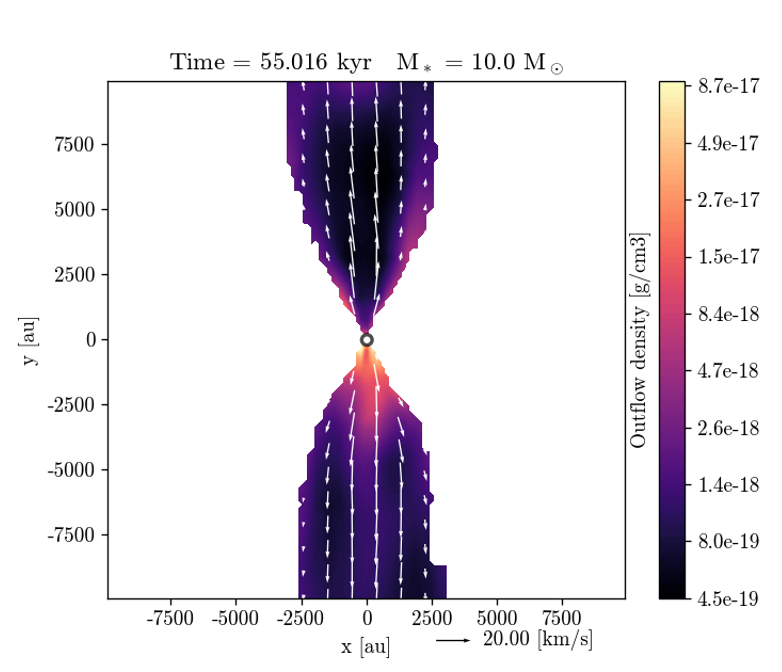} 
    \caption{Gas density slice of $20000$~AU in the outflow selection, perpendicular to the disk plane, run \textsc{NoTurb}, $M=10\Msol$. The gas outflow density corresponds to particle densities between ${\sim}10^5\, \mathrm{cm^{-3}}$ and ${\sim}10^7\, \mathrm{cm^{-3}}$.}
    \label{fig:rho_outflow}
\end{figure}

\begin{figure}
    \includegraphics[width=8cm]{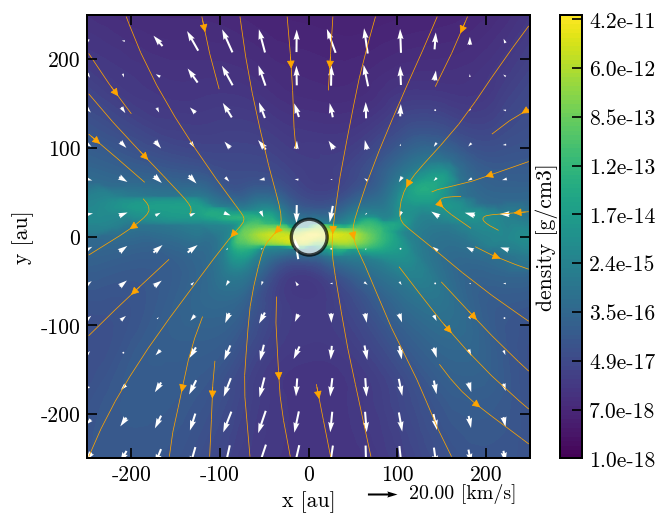}
    \caption{Gas density slice of $500$~AU perpendicular to the disk plane, run \textsc{NoTurb}, $M=10\Msol$. Velocity vectors and magnetic lines are overplotted. The gas density corresponds to particle densities between ${\sim}10^6\, \mathrm{cm^{-3}}$ and ${\sim}10^{13}\, \mathrm{cm^{-3}}$.}
    \label{fig:m0_rhomap_dx500}
\end{figure}

Above we have considered the total numerical radiative acceleration from our two radiative transfer modules, but it can be decomposed as the sum of the stellar radiative acceleration, treated with the M1 module, and the FLD radiative acceleration.
The latter corresponds to momentum transfer from dust-reprocessed (infrared-like) radiation, after stellar radiation (the main luminosity source in these simulations) has been absorbed. Figure~\ref{fig:fradfldm1} shows the ratio of FLD radiative acceleration to the gravitational acceleration (top panel) and to the M1 radiative acceleration (bottom panel).
The FLD acceleration also contributes to the outflow, since it dominates over the gravitational force.
Although its contribution is marginal compared to the direct stellar radiative force in the outflows here, it could play a more important role in the gas dynamics in the regions shielded from stellar radiation.
Indeed, the FLD acceleration is greater in the southern outflow, where density is higher (see the density slices displayed in Figs.~\ref{fig:rho_outflow} and \ref{fig:m0_rhomap_dx500}), due to more re-processed emission.
From the same figure, we observe regions of outflow density higher ($\rho > 10^{-18} \, \mathrm{g \, cm^{-3}}$) than in purely radiative outflows (see e.g., \citealt{rosen_unstable_2016}, \citealt{mignon-risse_new_2020}). As a consequence,  stellar radiation is absorbed and cannot contribute to the gas acceleration at large ($>10^4$~AU) distances when such a transient density region is present.
The ejection of optically-thick material is a common feature in our simulation, as we discuss below.

\subsubsection{Magnetic tower flow}

\begin{figure*}
\centering
    \includegraphics[width=8cm]{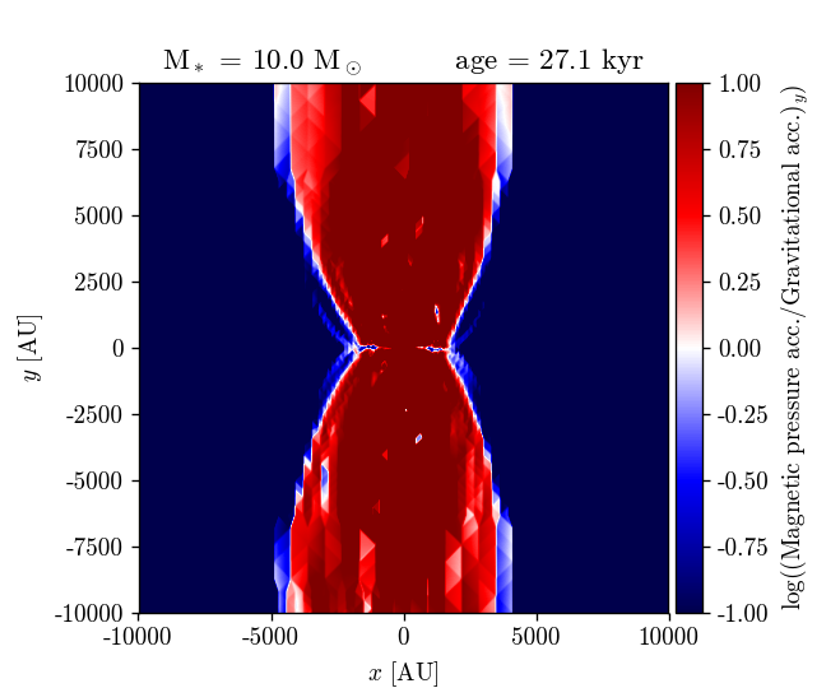}
    \includegraphics[width=9cm]{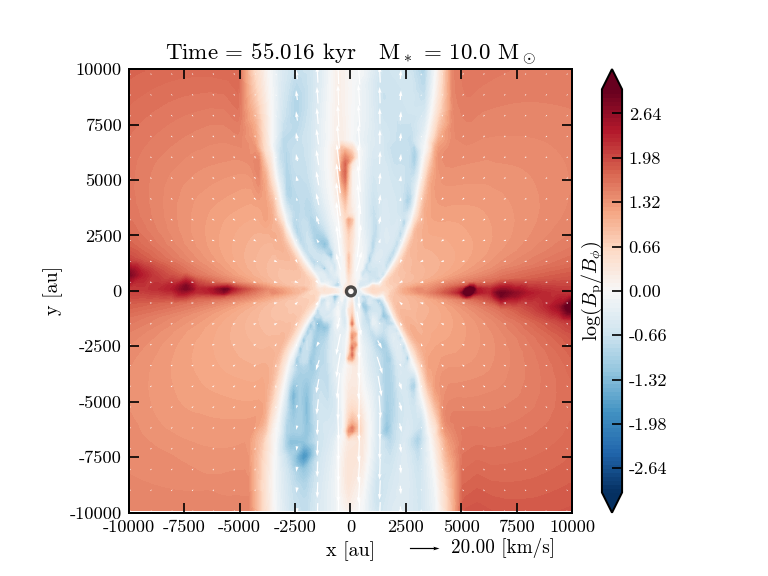}
    \caption{Left panel: ratio of the magnetic-pressure-gradient acceleration and the gravitational acceleration, in the vertical direction. Right panel: ratio of the poloidal and toroidal components of the magnetic field and velocity vectors overplotted. Slices of $20000$~AU perpendicular to the disk plane, when $M=10\Msol$, run \textsc{NoTurb}.}
    \label{fig:bpbphi}
\end{figure*}

\begin{figure}
\centering
    \includegraphics[width=9cm]{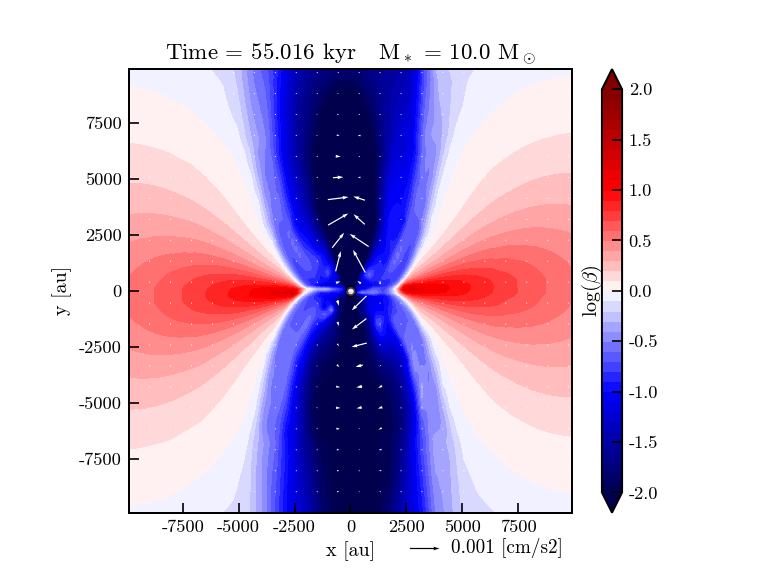}
    \caption{Plasma $\beta$ and Lorentz acceleration vectors overplotted. Slice of $20000$~AU perpendicular to the disk plane, when $M=10\Msol$, run \textsc{NoTurb}.}
    \label{fig:bpbphi2}
\end{figure}

Now, let us focus on the magnetic launching mechanism.
As shown in Fig.~\ref{fig:forcesratio}, the Lorentz force dominates the gas dynamics in the outflow.
It can be decomposed as the sum of a magnetic-pressure gradient force and a magnetic tension force.
While the former pushes the gas along the direction of stronger magnetic fields variations, giving rise to a magnetic tower flow, the latter impedes the bending of the field lines.
Left panel of Fig.~\ref{fig:bpbphi} shows the ratio of the magnetic-pressure-gradient force to the gravitational force in the direction perpendicular to the disk, computed from simulations outputs.
We only take the toroidal component of the magnetic field (in the frame of the sink), as it is the only one contributing to the gas dynamics in the poloidal direction \citep{spruit_magnetohydrodynamic_1996}.
This acceleration appears to dominate over gravity in all the outflow, by about one order of magnitude.
Therefore, the outflow in our simulation contains a magnetic tower flow (\citealt{lynden-bell_magnetic_1996}, \citealt{lynden-bell_why_2003}).
As shown in the right panel of Fig.~\ref{fig:bpbphi}, the toroidal component (blue) indeed dominates the outer zones of the outflow, while the poloidal component dominates close to the outflow axis.
In that respect, we obtain a similar outflow magnetic structure as many works in the literature (see e.g., \citealt{seifried_magnetic_2012}.
From the left panel of Fig.~\ref{fig:bpbphi} it can be seen that the tower flow launching region (i.e., close to the disk plane $y \,{\sim} \, 0$) is not restricted to the inner disk region because the disk radius in run \textsc{NoTurb} is ${\approx}100$~AU (see Paper I), while the region where this acceleration dominates over gravity (the red region) extends over more than $1000$~AU perpendicular to the outflow. This is consistent with the toroidal component of the magnetic field dominating beyond the disk outer radius (up to ${\approx}500$~AU, see Fig.~13 of Paper I). 
Actually, the tower flow develops on disk scales and widens later-on.
As in \cite{kato_magnetohydrodynamic_2004}, we find that the outflow itself is dominated by magnetic pressure ($\beta=P_\mathrm{th}/P_\mathrm{mag}<1$, where $P_\mathrm{th}$ and $P_\mathrm{mag}$ are the thermal and magnetic pressures, respectively, while the outflow edge corresponds to $\beta{\approx}1$), as displayed in Fig.~\ref{fig:bpbphi2}.
In addition to the possible thermal pressure gradient from the outer medium, collimation is enforced by the magnetic tension force when the field lines are sufficiently wound-up.
While we have emphasized the poloidal (i.e., pressure-driven) component of the Lorentz acceleration in the left panel of Fig.~\ref{fig:bpbphi}, there is a collimating component as well, as can be seen from the direction of the Lorentz acceleration vectors in Fig.~\ref{fig:bpbphi2}.
The tower grows vertically (i.e. the frontier between the outflow and the outer medium) as the field lines anchored on the disk rotate, and the tower vertical growth is predicted to occur at the disk rotation velocity \citep{lynden-bell_magnetic_1996}.
Indeed, looking at the evolution of the tower frontier position over $32$~kyr, we find a mean growth velocity of ${\approx}\,6\, \mathrm{km\, s^{-1}}$.
In the meantime, we reported a gas azimuthal velocity in the disk of ${\approx}\,5 \, \mathrm{km \, s^{-1}}$ at the outer radius.
This is consistent with \cite{lynden-bell_magnetic_1996}.

\subsubsection{Magneto-centrifugal outflow}

\begin{figure}
    \includegraphics[width=9cm]{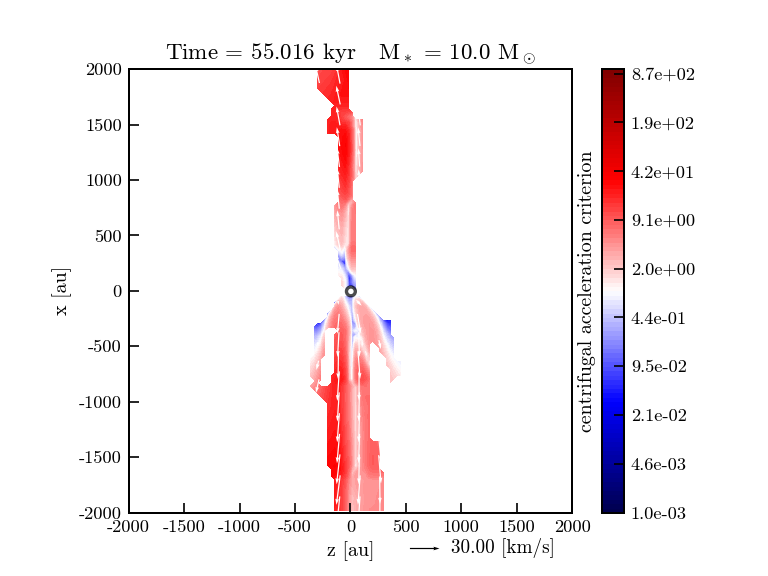}
    \caption{Criterion for centrifugal acceleration (Eq.~\ref{eq:centrif_crit}) from \cite{seifried_magnetic_2012} applied to a slice of $4000$~AU perpendicular to the disk plane. Run \textsc{NoTurb}, $M=10\Msol$. Red regions are consistent with centrifugal acceleration.}
    \label{fig:centrif_crit}
\end{figure}

Since the poloidal magnetic field component dominates close to the outflow axis and in the disk midplane (right panel of Fig.~\ref{fig:bpbphi}), we investigate whether the magneto-centrifugal process originally described by \cite{blandford_hydromagnetic_1982} is at work.
In this process, gas is centrifugally accelerated along field lines anchored in the disk and corotating with it.
Distinguishing centrifugal acceleration from a magnetic tower acceleration is a complicated task in such adaptive mesh refinement calculations, as underlined by \cite{seifried_magnetic_2012}.
In fact, the system is far from the ideal MHD, axisymmetric, stationary case and the criterion from \cite{blandford_hydromagnetic_1982} only applies to the disk surface.
They derived strict conditions in terms of magnetic field lines inclination to launch the flow centrifugally, but neglect disk thermal pressure which is obviously non-negligible in our calculation.
Moreover, analytical results rely on several invariants along the field lines (see e.g. \citealt{ogilvie_astrophysical_2016}), but it is difficult to trace the field line on which a gas particle has been centrifugally accelerated, back to the line foot point in the disk.
For that purpose, \cite{seifried_magnetic_2012} have derived a criterion to estimate whether centrifugal acceleration is taking place, based on grid-evaluated quantities.
They assume that $B_\phi=0$, so that the field lines corotate with the gas.
Since $B_\phi$ is never strictly equal to zero in our calculation, we apply this criterion only where $B_\mathrm{p}>B_\phi$.
Their idea is to determine, for a given point, the isocontour along which the effective gravity (accounting for the centrifugal force) is constant: it draws a line along which gas can freely move, regarding these forces.
Then they compare, in the $(r,z)-$plane (in cylindrical coordinates), the gas trajectory along this line to the field lines inclination, by computing the derivative $\partial z(r) / \partial r$ to $B_z/B_r$, where $z(r)$ is given by the isocontour equation (Eq.~16 of \citealt{seifried_magnetic_2012}).
Eventually, at any given point, centrifugal acceleration occurs if $\partial z(r) / \partial r$ is larger than the field line inclination, i.e.
\be
\log \left( \frac{r}{z} \frac{1}{\mathrm{G}M} \left( \frac{v_\phi^2}{r^2} (r^2+z^2)^{3/2} - \mathrm{G}M \right) \, \middle/ \, \left( \frac{B_z}{B_r} \right) \right) > 0,
\label{eq:centrif_crit}
\ee
where the numerator corresponds to $\partial z(r) / \partial r$.
We visualize this criterion in Fig.~\ref{fig:centrif_crit}: centrifugal acceleration occurs in red regions.
Hence, the zone close to the outflow axis, where we previously found $B_\mathrm{p}$ to dominate, is consistent with centrifugal acceleration.

\begin{figure}
    \includegraphics[width=9cm]{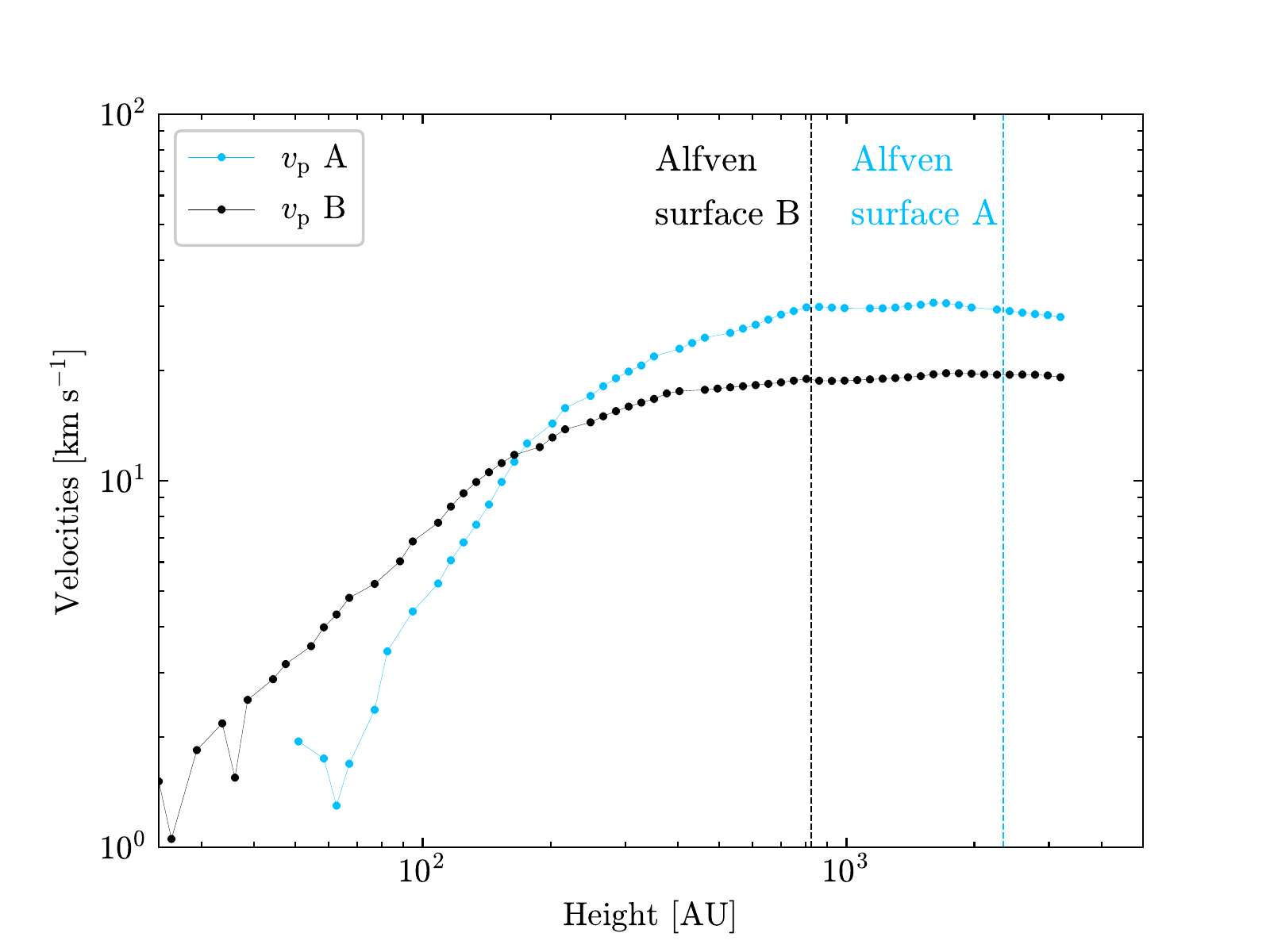}
    \caption{Poloidal velocity as a function of the distance to the sink (mainly in the vertical direction), in a cylindrical selection of cells with $r_\mathrm{cyl}<100$~AU. Negative radial velocities have been masked out. Velocities are obtained as averages over a distance bin. A and B label the northern and southern outflow, respectively, and the vertical lines indicate the positions where the poloidal velocity equals the Alfv\'en poloidal velocity (averaged over the same distance bin). Run \textsc{NoTurb}, $M=10\Msol$.}
    \label{fig:vpva}
\end{figure}

In the cold disk limit, gas is accelerated centrifugally from the disk surface to the Alfvén point, where the poloidal velocity equals the poloidal Alfvén speed.
We check this by visualizing these velocities as a function of the (mainly vertical) distance to the sink.
As shown in Paper I, $B_\mathrm{p}$ dominates for disk radii $\lesssim 50$~AU, hence the centrifugal mechanism may be at work below $50$~AU.
Therefore, we select cells at a cylindrical radius smaller than $100$~AU, so that their expected launching radius is a few tens of AU, consistently with the zone where the magnetic field is mainly poloidal within the disk.
Figure~\ref{fig:vpva} shows these velocities in the northern (A) and southern (B) outflows of run \textsc{NoTurb}, when $M=10\Msol$.
The poloidal velocity is found to increase when the distance to the sink is larger than $60-80$~AU.
Gas acceleration appears to take place up to the Alfvén point, in agreement with the theory (e.g., \citealt{spruit_magnetohydrodynamic_1996}).
As shown in the right panel of Fig.~\ref{fig:bpbphi}, even beyond the Alfvén surface ($\gtrsim 1000-2000$~AU), the poloidal component dominates, close to the outflow axis.
This feature is reminiscent of many studies including a magnetic tower flow (e.g. \citealt{kato_magnetohydrodynamic_2004}, \citealt{banerjee_massive_2007}, \citealt{seifried_magnetic_2012}, \citealt{kolligan_jets_2018}.
A plausible explanation for the generation of the poloidal component close to the axis (beyond the Alfvén surface) is the vertical inflation of the magnetic tower which develops the magnetic field poloidal component as it grows \citep{kato_magnetohydrodynamic_2004}.
Consistently, we find a nearly perfect alignment between the velocity vector and the magnetic field vector close to the outflow axis, while it is nearly perpendicular further away from the axis. This suggests that gas located near the axis is accelerated magneto-centrifugally.

The magneto-centrifugal mechanism is the best candidate for the fast outflows around young-stellar objects, hence we compare the highest velocities we obtain with theoretical predictions.
The terminal velocity $v_\infty$ is predicted to be (e.g. \citealt{pudritz_disk_2007})
\be
v_\infty {\simeq} \frac{r_\mathrm{c,A}}{r_\mathrm{c,0}} v_\mathrm{esc,0}\, {\simeq } \, 2-3 v_\mathrm{esc,0},
\ee
where $r_\mathrm{c,A}$ is the (cylindrical) Alfvén radius, $r_\mathrm{c,0}$ is the launching radius, so that $r_\mathrm{c,A}/r_\mathrm{c,0}$ is the lever arm and is typically $2-3$ \citep{pudritz_role_2019}, and $v_\mathrm{esc,0}$ is the escape velocity at the launching distance.
Magneto-centrifugal outflows have an onion-like velocity distribution, with the highest speed close to the axis corresponding to the gas initially close to the central object.
In our simulation, gas is launched at a vertical distance of $60-80$~AU from the sink (see also Fig.~\ref{fig:m0_rhomap_dx500}).
Hence, we infer a corresponding escape velocity of ${\approx}\,11 \mathrm{\, km \, s^{-1}}$, since $M=10\Msol$.
This leads to $v_\infty \, {\simeq} \, 22-33 \mathrm{\, km \, s^{-1}}$, which is of the same order as the fastest velocities we obtain at this time step, i.e. $v  {\sim}  32 \mathrm{  km \, s^{-1}}$ on one side of the disk and $v\,{\sim}\, 20 \mathrm{ km \,  s^{-1}}$ on the other side (Fig.~\ref{fig:vpva}). Hence, the magneto-centrifugal mechanism may be responsible for the outflow highest velocities, close to the axis, while the magnetic tower flow drives the wider-angle and slower component of the outflow. Moreover, the wide-angle gas is unlikely related to magneto-centrifugal acceleration because it can be located more than $2000$~AU away from the axis (see Fig.~\ref{fig:rho_outflow}), which is inconsistent with a launching from a $100$~AU disk with a lever arm of $2-3$ as predicted by the theory.

Let us note that the highest velocity in each lobe shows fluctuations between these two values.
These small velocity differences suggest that this mechanism may be either transient in our simulation (the radiative acceleration being able to accelerate the gas to $v\, {\sim}\,20 \mathrm{\, km \, s^{-1}}$) or not symmetric with respect to the disk plane (as can be seen in Fig.~\ref{fig:rho_outflow}).
This north-south asymmetry in the ejection may arise from the asymmetry in the streamers.
These channels feeding the disk are not located in the disk plane (more details in Paper I), hence part of the outflowing gas may inherit from this asymmetry.

Let us also recall that the magneto-centrifugal mechanism taps in the gravitational energy, as can be seen from the relation above between the outflow terminal velocity and its initial escape velocity. Hence, a launching from the disk (${\sim}20$~AU for the disk inner edge) instead of ${\sim}100$~AU above it would result in an initial escape velocity (and therefore a terminal velocity) more than twice larger.
Overall, there are several clues indicating the presence of a magneto-centrifugal jet in our simulation.

\subsection{Influence of a turbulent medium: runs \textsc{SupA}, \textsc{SupAS}, \textsc{SubA}}

\begin{table}
\caption{Simulations outcomes regarding the outflow launching.}
\label{table:results}
\centering 
\begin{tabular}{l | l l | l l l} 
    \hline \hline
	 Model & $t_\mathrm{out}$ [kyr]  &  $M_{\star}(t_\mathrm{out})$ [$\mathrm{M_\odot}$] & Outflow  \\ \hline
	\textsc{NoTurb}    & $36.0$ & $3.7$ & bipolar \\ \hline 
    \textsc{SupA}	  & $56.4$ & $6.6$ & bipolar \\ \hline
	\textsc{SupAS}	  & $66.2$ & $5.1$ & unipolar (transient)  \\ \hline
	\textsc{SubA} & $39.1$ & $3.8$ & bipolar \\ \hline
\end{tabular}
{\raggedright \footnotesize{\textbf{Notes.} $t_\mathrm{out}$ (kyr) denotes the time when sustained outflows appear, $M_\star(t_\mathrm{out})$ ($\Msol$) is the primary sink mass at this time.} \par}
\end{table}

\begin{figure*}
    \includegraphics[width=18cm]{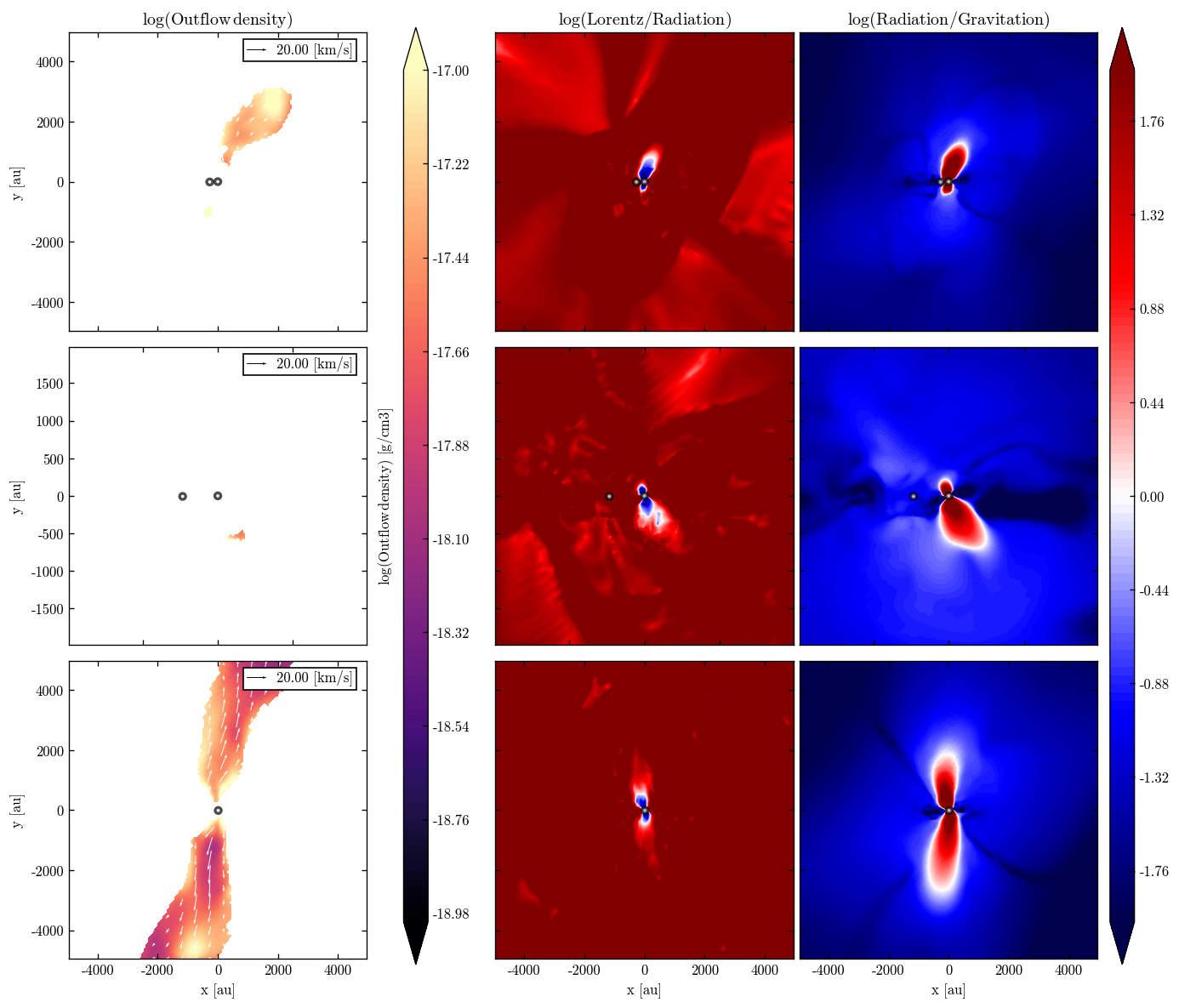}
    \caption{Slices perpendicular to the disk plane. Left column: density slice in the outflow selection. Middle column: ratio of the Lorentz acceleration to the radiative acceleration. Right column: ratio of the radiative acceleration to the gravitational force. From top to bottom: run \textsc{SupA} (super-Alfv\'enic, subsonic turbulence, $10000$~AU, $t=67.0$~kyr, $M=8.2\Msol$, $L=1.4\times 10^4\Lsol$), run \textsc{SupAS} (super-Alfv\'enic, supersonic turbulence, $4000$~AU, $t=72.6$~kyr, $M=5.6\Msol$, $L=8\times 10^3\Lsol$) and run \textsc{SubA} (sub-Alfv\'enic, subsonic turbulence, $10000$~AU, $t=61.1$~kyr, $M=9.6\Msol$, $L=1.7\times 10^4\Lsol$). The gas densities in the left column correspond to particle densities between ${\sim}10^4 \, \mathrm{cm^{-3}}$ and ${\sim}10^{6} \, \mathrm{cm^{-3}}$.} 
    \label{fig:forcesturb}
\end{figure*}

We now focus on the outflows in the three other runs.
Figure.~\ref{fig:forcesturb} shows density slices in the outflow selection (left panel), the ratio between the Lorentz and the radiative accelerations (middle panel) and the ratio between the radiative and the gravitational accelerations (right panel).
We recall that the Lorentz acceleration encapsulates the magnetic pressure gradient acceleration.
Outflows form at $t{\sim}30$~kyr in the sub-Alfv\'enic runs, \textsc{NoTurb} and \textsc{SubA}.
Meanwhile, their launching occurs at $t=56$~kyr in run \textsc{SupA} and ${\sim}66$~kyr in run \textsc{SupAS} (see Table~\ref{table:results}).

The inclusion of a non-coherent initial velocity distribution in our turbulent runs should perturb the magnetic field coherence, impeding the launching of the outflow.
As shown in Fig.~13 of paper I, ${\sim}22$~kyr after sink formation a strong toroidal magnetic field has built up, but no outflow has been launched yet in runs \textsc{SupA} and \textsc{SupAS}.
Indeed, the density structure formed by the combined effect of infall and turbulent motions is a filament-like structure of a few thousands AU almost perpendicular to the disk plane, which carries an additional ram pressure to be overcome by the outflow, no matter its origin.

Magnetic and radiative forces have different natures.
On the one hand, magnetic outflow launching is a long-term process and can be prevented, e.g. by the orbital motions in a binary system \citep{peters_interplay_2011}.
On the other hand, the launching (close to the star) of radiative outflows is isotropic and depends mostly on the density distribution, via the optical depth. Its launching and propagation depend on the environment, so one can expect transient and smaller radiative outflows in a turbulent medium, unless radiation can find its way out and accelerate gas instantaneously.
Without magnetic fields, \cite{rosen_massive-star_2019} have found that infalling filaments of gas are self-shielded against radiation and form a network of dense filaments and optically-thin channels centered on the massive star.

In the present study, with magnetic fields and super-Alfv\'enic turbulence (run \textsc{SupAS}), gravity is diluted and material gently falls via thermally-supported ($\beta>1$) streamers on a moderately-magnetized complex structure of ${\sim}1000$~AU squared (see Fig.~2 of Paper I). 
At that time, a secondary star-disk system has formed.
As a consequence, we observe two failed attempts of launching outflow, as dense gas passes through it.
These occur when the secondary sink is closer to the apastron.
Eventually, the monopolar outflow launches, and survives for ${\sim}\, 3$~kyr before it becomes difficult to characterize it as an outflow, since it has been perturbed by the environment motions and no gas is newly ejected from the basis.
A similar process occurs in run \textsc{SupA}.
While the ram pressure is lower than in run \textsc{SupAS} and consequently, an outflow successfully developed, the formation of a secondary sink at about the same time has progressively displaced the center of mass of the system.
The primary sink disk moves on a ${\sim}350-600$~AU orbit and the outflow is broadened, from the basis, consequently.
Nonetheless, it is sustained until the end of the run, oppositely to run \textsc{SupAS}.
As mentioned previously, the orbit is eccentric.
When the primary sink approaches the apastron, it stays longer in the same area and has more time to accelerate the gas radiatively.
Finally, despite the turbulent support, the sub-Alfv\'enic run \textsc{SubA} has no difficulties launching the outflows at about the same time as in the fiducial run, because the initial magnetic field is stronger.
The toroidal magnetic field reaches similar values as in the less-magnetized, non-turbulent run \textsc{NoTurb} ($>0.1$~G).
The magnetic tower develops at about the same speed as in run \textsc{NoTurb} (middle panel of Fig.~\ref{fig:outflowsppt}).
The presence of a turbulent velocity field contributes to the "north-south" asymmetry. 
The bipolar outflows, which are not strictly identical in run \textsc{NoTurb}, are even more distinguishable in terms of extent or orientation here (middle and right panel of Fig.~\ref{fig:t_th}).
Hence, turbulence provides an additional mechanism to break the symmetry between bipolar outflows and can even suppress them.

\begin{figure}
    \includegraphics[width=9cm]{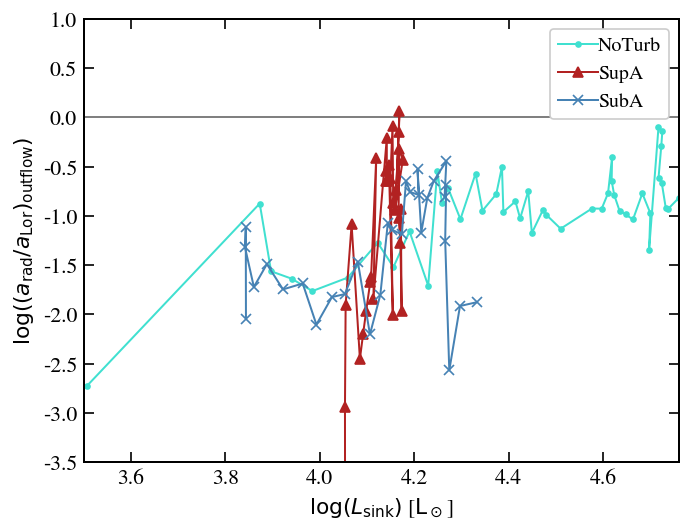}
    \caption{Ratio between the radiative and Lorentz accelerations (both integrated over the outflow volume), as a function of the sink luminosity.}
    \label{fig:t_fradlor}
\end{figure}

Figure~\ref{fig:forcesturb} shows that once the outflows are launched in runs \textsc{SupA} and \textsc{SupAS}, the local relative contribution from radiative acceleration to the total acceleration is larger than in the fiducial case. First, by delaying the launching, the central star has time to reach slightly higher masses (hence, luminosities). Second, the magnetic field is less organized than in the non-turbulent case, thus the component of the Lorentz force contributing to the outflow is smaller.

Let us compare the two accelerations in the outflow as a function of time.
Figure~\ref{fig:t_fradlor} shows the ratio between the radiative acceleration and the Lorentz acceleration, both integrated over the outflow volume, as a function of the primary sink luminosity. In our simulations, the sink luminosity is an increasing, monotonic function of time.
This figure shows that the Lorentz acceleration is significantly greater (two orders of magnitude) than the radiative acceleration at the time when the outflow forms.
In run \textsc{SupA}, the ratio approaches one. This is due to the outflow having formed later than in the other runs, so the outflow is smaller and radiative acceleration is efficient. We observe that, even for a luminosity larger than $10^4 \mathrm{L_\odot}$, the Lorentz acceleration dominates in the outflow. 

As in run \textsc{NoTurb}, we find the poloidal component of the magnetic field dominating the toroidal component close to the outflow axis and in the disk plane (see Paper I) in runs \textsc{SupA} and \textsc{SubA}. This suggests that the magneto-centrifugal mechanism could be at play, in addition to the Lorentz acceleration.

To conclude, turbulence delays the outflows but does not change their nature: we still obtain magnetic outflows, although the local relative contribution from radiative acceleration is larger than without turbulence.

\subsection{A channel for radiation?}
\label{sec:outflowchannel}

The magnetic outflows develop at a smaller stellar mass ($M\,{\approx}\,4-7\Msol$, see Table~\ref{table:results}) than what is found in RHD simulations regarding radiative outflows ($M>10\Msol$, see e.g. \citealt{kuiper_stability_2012}, \citealt{mignon-risse_new_2020}).
Hence, they could act as a channel of radiation to propagate, as proposed by \cite{krumholz_how_2005} for protostellar outflows. \cite{banerjee_massive_2007} proposed the same mechanism for tower flows, but their calculation did not include radiative transfer.
Despite the regular presence of optically-thick gas in the outflow, most of the outflow volume is optically-thin.
To assess the effect of the radiative force, we compare the outflow extent between the \textsc{NoTurb} run and one including the FLD method rather than the hybrid method (that we will call the \textsc{NoTurbFLD} run, see the Appendix~\ref{fig:app_fldhyout}).
When the central star is ${\sim}5\Msol$, the outflow extends over more than ${\sim}4500$~AU in the \textsc{NoTurb} run while it extends over $3000$~AU in run \textsc{NoTurbFLD} (see Fig.~\ref{fig:app_fldhyout}).
Moreover, the outflow appears more symmetric (axisymmetric and north-south) in the \textsc{NoTurb} run than in the \textsc{NoTurbFLD} run, indicating that the radiative force stabilizes the outflow structure.
To sum up, the outflow does appear as a channel for radiation to escape. Radiative acceleration participates to the gas acceleration, more than in the FLD case, as we find that the highest gas velocity is $25\%$ smaller in run \textsc{NoTurbFLD} than in \textsc{NoTurb} (see Appendix~\ref{fig:app_fldhyout}).

\subsection{Outflow properties}

\subsubsection{Outflow mass}
\label{sec:outflowmass}

\begin{figure}
\centering	
    \includegraphics[width=8cm]{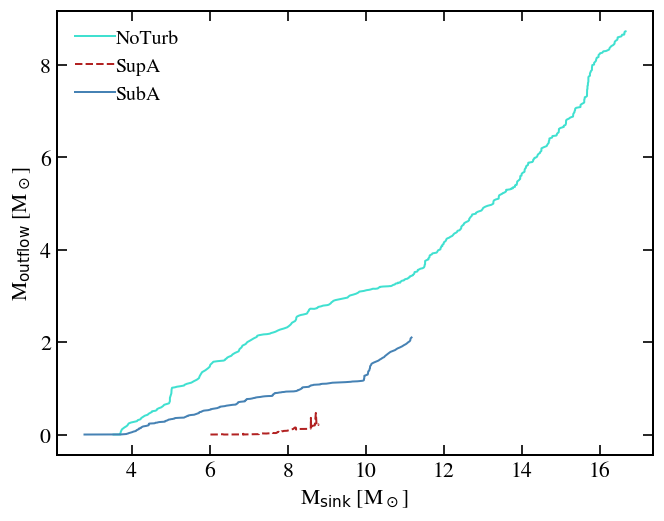}
    \caption{Outflow mass as a function of the sink mass.}
    \label{fig:msmo}
\end{figure}

Figure~\ref{fig:msmo} shows the outflow mass as a function of the sink mass.
It generally increases with time and has values $1-8\Msol$ in sub-Alfv\'enic runs and subsolar masses in run \textsc{SupA} during the epoch covered. 
While it appears to be variable in run \textsc{SupA}, it only increases in sub-Alfv\'enic runs, and more rapidly in the non-turbulent run \textsc{NoTurb}. 
We note that step 4 of our outflow definition (removing cells far from the outflow geometric center and close to the secondary sink) is required to get relevant measurements of outflow mass in run \textsc{SupA}. Without this criterion, the outflow mass is larger by one order of magnitude because of the dense gas gravitationally bound to the secondary sink being loosely accounted for.
Considering their mass and dynamical time (i.e. timescale of existence), we obtain a mean ejection rate of ${\sim}\,2\times 10^{-4}\, \,\Msolyr$ in run \textsc{NoTurb}, 
${\sim}\,5\times 10^{-5}\, \, \Msolyr$ in run \textsc{SubA} and ${\sim}\,10^{-5}\, \,\Msolyr$ in run \textsc{SupA}.

It can be noted that around $11\Msol$ and $10\Msol$ there is a small change of slope in the outflow mass evolution, in runs \textsc{NoTurb} and \textsc{SubA}, respectively.
Interestingly, radiative outflows are reported to occur at about this mass, in radiation-hydrodynamical simulations  (\citealt{kuiper_stability_2012}, \citealt{mignon-risse_new_2020} with the same Pre-Main Sequence track as here, i.e. taken from \citealt{kuiper_simultaneous_2013}).
Hence, the change of slope, and more specifically the increase in the outflow mass to sink mass ratio may be linked to the increasing radiative force.
An argument in that regard comes from the comparion with C21.
They measure an outflow mass of ${\sim}2\Msol$ when the sink is ${\sim}8\Msol$, which is similar to what is obtained here.
Since the main difference between our runs comes from the radiative transfer method used, and the Flux-Limited Diffusion underestimates the direct stellar force compared to the hybrid method, this change of slope appearing at the stellar mass of ${\sim}20\Msol$ instead of ${\sim}10\Msol$ is consistent with a radiative force origin.

\subsubsection{Momentum rate}
\label{sec:outflmomrate}

\begin{figure*}
\centering
    \includegraphics[width=6cm]{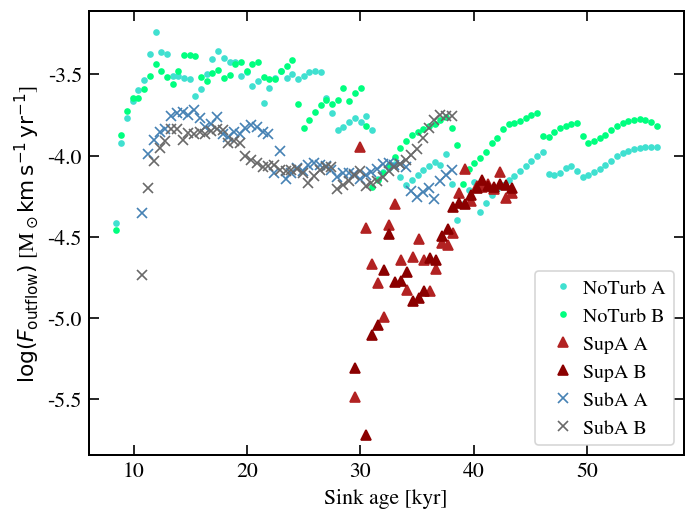}
    \includegraphics[width=6cm]{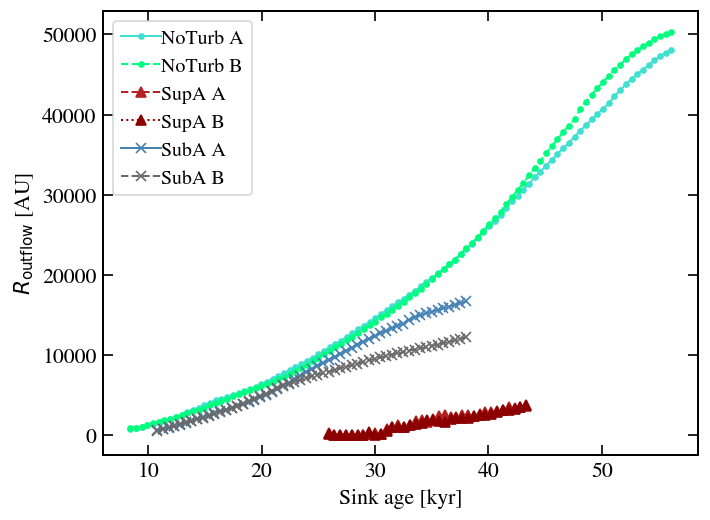}
    \includegraphics[width=6cm]{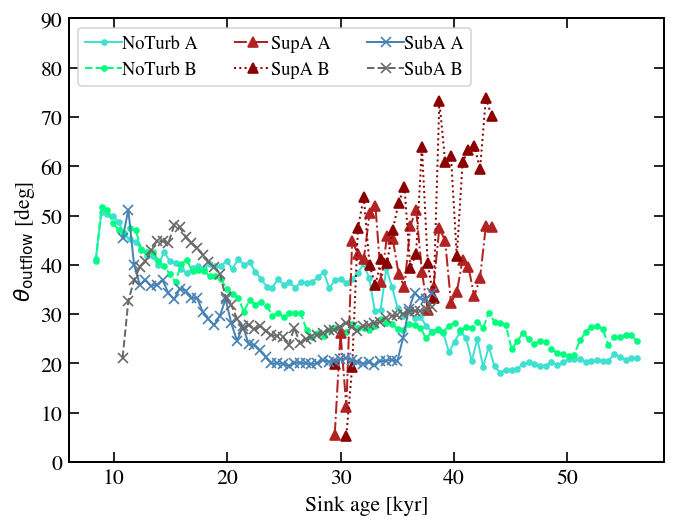}
    \caption{Outflows properties as a function of the primary sink age: momentum rate (left), maximal outflow radius (middle), opening angle (right) and angle between the outflow and the large-scale magnetic field (bottom-right), in runs \textsc{NoTurb}, \textsc{SupA}, \textsc{SubA}. Forces and opening angles of outflows composed of less than $50$ cells are not displayed. Outflows come by pair in these runs, so they are individually labeled as A and B. Values are averaged over $0.5$~kyr (smaller than the orbital timescale in run \textsc{SupA}).}
    \label{fig:outflowsppt}
\end{figure*}

\begin{figure*}
\centering
    \includegraphics[width=6cm]{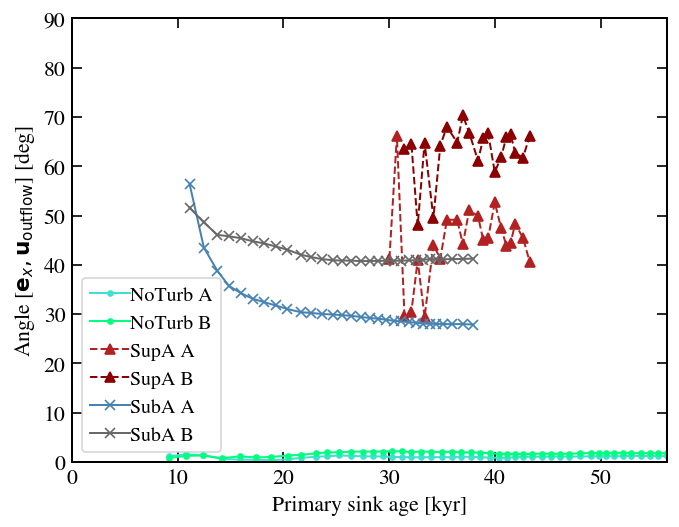}
    \includegraphics[width=6cm]{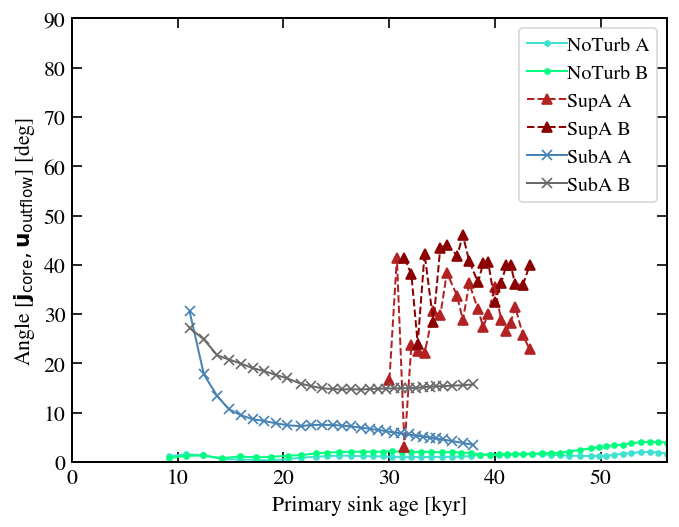}
    \includegraphics[width=6cm]{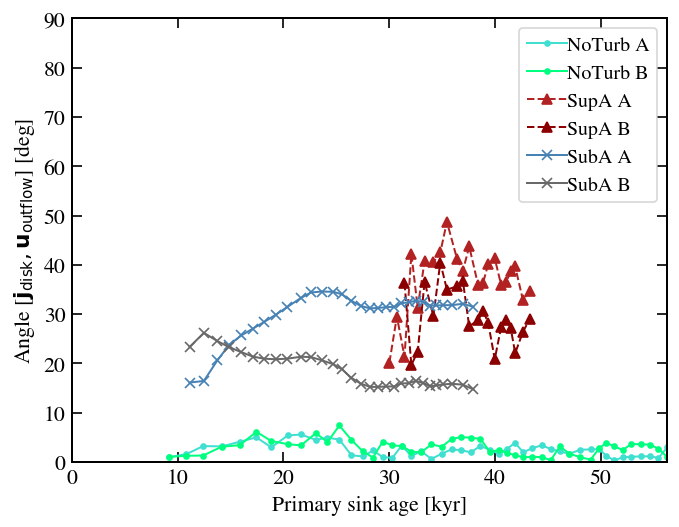}
    \caption{Angle between the outflow and the large-scale magnetic field (left), the core-scale angular momentum (middle) and the disk (right), respectively, in runs \textsc{NoTurb}, \textsc{SupA}, \textsc{SubA}. Values obtained for composed of less than $50$ cells are not displayed. Outflows come by pair in these runs, so they are individually labeled as A and B.}
    \label{fig:t_th}
\end{figure*}

Left panel of Fig.~\ref{fig:outflowsppt} displays the outflow momentum transfer rate (also called outflow force) computed from Eq.~\ref{eq:foutflow}, each point corresponding to an outflow (either northern, labeled "A" or southern, "B") at a given time step.
For runs \textsc{NoTurb} and \textsc{SubA}, we have $F_\mathrm{outflow}$ of the order of $10^{-4} \Msol \, \mathrm{ km \, s^{-1} \, yr^{-1}}$ and a dispersion of less than one order of magnitude.
We observe more dispersion at the beginning of run \textsc{SupA}, then the evolution is similar with an overall increasing force with time. By the end of run \textsc{SupA}, the outflow force reaches similar values as in runs \textsc{NoTurb} and \textsc{SubA} with ${\sim} 10^{-4} \Msol \, \mathrm{ km \, s^{-1} \, yr^{-1}}$.  
These are consistent with the aforementioned numerical work of \cite{seifried_magnetic_2012}.

\subsubsection{Opening angles}
\label{sec:outflopang}

Close to the star, the outflow shape resembles a conical shape before collimation occurs ($\lesssim 2000$~AU) and extends the outflow in an elliptic shape.
In Sect.~\ref{sec:diskdef} we have presented our method to compute the outflow opening angle (see also Fig.~\ref{fig:openingscheme}).
We have adopted a method adapted to the elliptic shape of the outflows we observe, which is similar to \cite{offner_radiation-hydrodynamic_2011}.

The right panel of Fig.~\ref{fig:outflowsppt} shows $\theta_\mathrm{outflow}$ as a function of the sink age.
We mentioned above that the outflow launched was quite similar between runs \textsc{NoTurb} and \textsc{SubA}.
Consequently, the values and evolution of the opening angle are, to first order, similar.
During a first phase (a few kyr), the outflow broadens so $\theta_\mathrm{outflow}$ increases, then (after a sink age of roughly $9$~kyr in run \textsc{NoTurb}, $11$~kyr and $16$~kyr in run \textsc{SubA}) the base of the outflow becomes nearly stationary but the outflow propagates, hence the opening angle decreases.
During this second phase, the angle has values of $20$ to $40\deg$ which are north-south asymmetric.
Finally, it tends toward $20-25\deg$.
The outflow re-collimates, which is partly due to the toroidal component of the magnetic fields (Fig.~\ref{fig:bpbphi2}) and possibly to the pressure from the outer medium, in addition to the aforementioned geometrical effect.
In run \textsc{SupA}, the measurement of the opening angle is greatly affected by the orbital motions of the sink because the orbital separation (as large as $600$~AU) is not negligible with respect to the outflow extent (${\sim}2000$~AU, middle panel of Fig.~\ref{fig:outflowsppt}), and an orbital velocity similar to the tower growth speed, by definition (Sect.~\ref{sec:outflows}).
Since both velocities and spatial extents are of the same order, the outflow geometry becomes complex.
Hence, the opening angle in run \textsc{SupA} is not comparable to a single observation.
If any, it shows that the stellar motions in a turbulent medium, or a multiple stellar system, will strongly affect this type of geometrical measurements. 
Overall, we obtain opening angles varying between $30\deg$ and $70\deg$ and between the north and south outflow.
The orbital motion seems to have played a dominant role in the outflow broadening.
We will focus on the outflow orientation in the following section.

\subsection{Alignment with magnetic fields, core-scale angular momentum and disk}
\label{sec:alignoutflow1}

Low- and high-mass pre-stellar cores are threaded by magnetic fields, but their exact role is not clear yet.
Since disk-mediated accretion is observed in the low-mass regime (e.g., \citealt{pety_plateau_2006}), and now in the high-mass regime as well (see e.g. \citealt{cesaroni_chasing_2017}), and disks are required to launch MHD outflows (supported by e.g., \citealt{hirota_disk-driven_2017}), studying the alignment between outfows and magnetic fields should provide insights onto their exact role during (massive) star formation. Furthermore, magnetic outflows are expected to be launched perpendicular to the disk. In the following we study the misalignment between outflows and magnetic fields, angular momentum (on core scale) and disk.
Figure~\ref{fig:t_th} shows the angle formed by the outflow geometric center vector with respect to the $x-$axis (corresponding to the initial magnetic field orientation, left panel), with respect to the core-scale angular momentum vector (middle panel) and with respect to the disk normal vector (right panel), as a function of time.

In run \textsc{NoTurb}, we find a nearly perfect alignment between between the outflows and the magnetic fields, the core-scale angular momentum, and the disk normal.
Several factors have broken the north-south symmetry as well as the axisymmetry (which could increase the outflow-disk misalignment), still the misalignment is smaller than $10 \deg$ in each case and the bipolar outflows show similar misalignment angles.

Let us now study the misalignments in the turbulent runs.
In run \textsc{SupA}, the bipolar outflows are not symmetric and there is no clear trend toward an alignment with the large-scale magnetic fields.
Most of the time, the outflows align within less than $40\deg$ with the disk normal and with the core-scale angular momentum.

In run \textsc{SubA}, the angles between the outflows and both magnetic fields, and core-scale angular momentum decrease with time (but never reach a perfect alignment), suggesting a preference for outflow-angular momentum and outflow-magnetic fields alignments on large scales.
This is naively expected since magnetic outflows are related to organized field lines twisted by rotation.
Here is another possible interpretation, based on the presence of streamers (dense filaments) perpendicular to the magnetic fields (see Fig.~2 of Paper I), randomly oriented with respect to the disk.
Streamers either may put forbidden directions for the outflows by opposing a strong ram pressure, and these forbidden directions are $90\deg$ oriented with respect to the magnetic fields, or bring angular momentum and contribute to twisting the field lines.
By preventing outflow launching along these directions, the outflow center of mass is shifted toward a location closer to the magnetic fields axis.
This trend is not visible in run \textsc{SupA}, where the angles do not show any clear evolution other than periodic variations on orbital timescales.

Finally, a few words on the short-lived (${\approx}\,3$~kyr) monopolar outflow in run \textsc{SupAS}.
It develops nearly-perpendicular to the disk (with a disk-magnetic field misalignment of ${\sim}90\deg$, see Paper I).
It shows that, indeed, a disk perpendicular oriented to the core-scale magnetic fields has trouble launching outflows but this is possible though \citep{joos_influence_2013}.
It can also occur on smaller scales than those covered in this study, especially in the case of magneto-centrifugal jets where the highest velocity component comes from the disk inner radius.

To sum up, in the four runs, the outflow orientation appears to be mainly set by the disk orientation, which depends on the initial angular momentum.
Nonetheless, it never corresponds to a strict perpendicular angle with the disk, and is larger in the super-Alfv\'enic run than in the sub-Alfv\'enic run.
As the outflow grows, it tends to align with the core-scale magnetic fields and angular momentum when turbulence is sub-Alfv\'enic.
Overall, the alignment with the disk normal and with the core-scale angular momentum (which is linked to the disk normal, as shown in Paper I) are better than with magnetic fields.

\section{Comparison of the outflow properties with observational constraints}
\label{sec:outflowsobs}

In the following, we compare the outflows properties to several observational studies based on low- and high-mass protostars statistical samples.
When comparing to low-mass objects, we implicitly assume a continuity in the outflow launching mechanism from low- to high-mass protostars, as pointed out by many studies (i.e., \citealt{cabrit_co_1992}, \citealt{bally_protostellar_2016}).

\subsection{Outflow velocity, mass, dynamical time, ejection rate}

As mentioned in the previous section, the outflows in our simulations are dominated by MHD processes while radiation can participate to the acceleration. 

Before comparing the outcomes of our simulations with observational values, let us precise that some of these observable quantities are often plotted against the stellar luminosity (see e.g. \citealt{lada_cold_1985}).
The luminosity does not only stand as a tracer of the evolutionary stage.
Since high-mass protostars are expected to have higher accretion rates than their low-mass counterparts \citep{motte_high-mass_2018}, the luminosity has often been used as a proxy for the accretion rate \citep{wu_study_2004}.
This is of main interest here, since MHD disk outflows are powered by the gravitational energy from accretion, with a predicted ratio of mass outflow rate to mass accretion rate ${\sim}0.1$ (see \citealt{pudritz_role_2019} and references therein).
\cite{matsushita_massive_2017} obtain a ratio ${\gtrsim}0.2$ which can approach unity when the core initial magnetic energy is comparable to the gravitational energy.
Finally, we will refer to a mean accretion/ejection rate by run, rather than an instantaneous rate as it can vary on more than one order of magnitude from one timestep to the other (see Paper I).

First, as shown in the left panel of Fig.~\ref{fig:centrif_crit}, the maximal outflow velocity $v_\mathrm{max}$ in run \textsc{NoTurb} is ${\simeq}20 \mathrm{\, km \, s^{-1}}$ for one outflow lobe and ${\simeq}32 \mathrm{ \, km \, s^{-1}}$ for the other, at the time when the central star is $10 \Msol$.
This velocity is expected to gently increase with the squared root of the sink mass for magnetic outflows; after a sudden increase phase (until $M_\mathrm{sink} {\sim}6 \Msol$), we find $v_\mathrm{max}/\sqrt{M_\mathrm{sink}}$ to remain constant within ${\sim}20\%$.
We compare the previous values with those obtained by \cite{nony_episodic_2020} on the most massive core (${\sim}102 \Msol$) of their $1-100\Msol$ sample (in the W43-MM1 protocluster). On this sample, they obtained a median velocity of $47 \mathrm{\, km \, s^{-1}}$.
The most massive core exhibits a monopolar outflow with a maximal velocity of $34\pm 2 \mathrm{km \, s^{-1}}$ and $10000 \pm 1000$~AU length, which agrees well with one of the two outflow lobes in run \textsc{NoTurb} (when the central star is $10 \Msol$).
Interestingly, while we have attributed the monopolar nature of the outflow in run \textsc{SupAS} to the ram pressure of the turbulent gas, this occurence in W43-MM1 could be due to an inflow of material according to \cite{nony_episodic_2020}.

Let us first present the observational results regarding outflow masses before comparing with our study. 
\cite{wu_study_2004} built a statistical study of $391$ high-velocity outflows, covering several evolutionary stages.
For $L>10^3 \Lsol$ objects, they obtain outflow masses of a few solar masses up to $10^2\Msol$ with averaged dynamical times of $100$~kyr.
This is consistent with the study of \cite{beuther_massive_2002}, focused on the CO $J=2-1$ emission towards $26$ massive star-forming regions.
They obtain outflow masses of typically $M_\mathrm{outflow} {\sim} 0.1 (\mc/\Msol)^{0.8} \Msol$ (where $\mc$ is the core mass) and dynamical time scales of the order of the core free-fall time.
In the sample of $11$ massive star-forming regions of \cite{wu_co_2005},
the outflow mass is found to be between a few solar masses too, while the maximal mass is $60\Msol$ and averaged dynamical timescales of $20$~kyr.
Similarly, \cite{zhang_search_2005} extract a mean outflow mass of $20.6\Msol$ and a median of $15.6\Msol$ from a sample of $69$ sources with luminosities $10^{2-5} \Lsol$.

The upper-mass limits of $60-100\Msol$ are significantly larger than what we obtain, as well as the values of $15.6-20.6\Msol$ of \cite{zhang_search_2005}, although the latter values might be reached at later times in our study (this would occur at $M{\sim}21\Msol$ in run \textsc{NoTurb}, extrapolating on the results presented in Fig.~\ref{fig:msmo}).
The outflow mass presented in \cite{beuther_massive_2002} for a core mass similar to ours ($\mc = 100\Msol$, corresponding to $L>10^3\Lsol$ from their Fig.~4) is ${\sim}4\Msol$ (see the relation above), which is consistent with our results for sub-Alfv\'enic runs (Fig.~\ref{fig:msmo}), and possibly for run \textsc{SupA} at later times.
All these studies agree on typical accretion rates of a few $10^{-4}\,  \Msolyr$, similar to those presented in Paper I.
Hence, regarding the outflow mass, our outflows are consistent with observational constraints.

On the one hand, the outflow ejection rate is consistent with observations of high-mass cores and luminous ($>10^2\Lsol$) protostellar objects.
On the other hand, the outflow mass agrees when the core mass is $100\Msol$ \citep{beuther_massive_2002}, and is smaller than for more massive cores.
This discrepancy can be explained by our initial conditions corresponding to the low-mass limit of massive cores.

\subsection{Outflow momentum rate}

Let us compare the results presented in Sec.~\ref{sec:outflmomrate} with the current observational constraints (observed in CO), for $L>10^3\Lsol$ objects.
Indeed, the pioneer study of \cite{lada_cold_1985} has shown a general trend between the outflow force and the stellar luminosity of $F_\mathrm{outflow} {\sim} 10^2 L/\rmc$ from $1\Lsol$ to $10^5\Lsol$, suggesting a common outflow mechanism for low- and high-mass protostars, which is likely a magnetic mechanism.
Hence, let us determine whether our outflow forces are consistent with this trend, with up-to-date outflow samples.
In the statistical analysis of \cite{wu_study_2004} towards high-velocity outflows, $10^2 \Msol$ is the lowest core mass of the sample and gives $F_\mathrm{CO} = 10^{-3}\Msol \, \mathrm{km \, s^{-1} \, yr^{-1}}$.
Towards $11$ massive star-forming regions, \cite{wu_co_2005} found values between ${\sim}10^{-3}\Msol \, \mathrm{km \, s^{-1} \, yr^{-1}}$ and $2 \times 10^{-1}\Msol \, \mathrm{km \, s^{-1} \, yr^{-1}}$ (for $L>10^3\Lsol$ protostars).
Including the measurements from \cite{beuther_massive_2002}, \cite{zhang_search_2005} obtain outflow forces of $10^{-4}-10^{-2} \Msol \, \mathrm{km \, s^{-1} \, yr^{-1}}$.
Hence, the outflow momentum rate we obtain is consistent with the lower values mentioned above, that is $10^{-4}\Msol \, \mathrm{km \, s^{-1} \, yr^{-1}}$.
We note that the uncertainty is almost two orders of magnitude on the values of \cite{wu_co_2005} though.
Further observational campaigns are required to put stronger constraints on the outflow force.

\subsection{Opening angles}

Collimated outflows are observed around O- and B-type protostars \citep{arce_molecular_2007}, but several studies point toward less collimated outflows in the high-mass regime than in the low-mass regime (see e.g., \citealt{beuther_massive_2002}, \citealt{wu_study_2004}).
Opening angles between $17\deg$ and $25\deg$ (that is, a good collimation) have been reported in the massive protostellar sources
IRAS 20126+4104 \citep{moscadelli_water_2005} and IRAS 16547-4247 
\citep{rodriguez_high_2005}, but likely originate from a magneto-centrifugal jet given the velocities involved ($34$ to $112 \, \mathrm{km \, s^{-1}}$ for IRAS 20126+4104).

The outflow morphology below ${\sim}2000$~AU in runs \textsc{NoTurb} and \textsc{SubA} roughly fits a conical shape.
The outflow growth, while keeping this shape, lasts a few kyr.
This epoch corresponds to the highest values of $\theta_\mathrm{outflow}$ measured, with $\theta_\mathrm{outflow}{\approx}30-60\deg$ until then.
For comparison, \cite{pety_plateau_2006} (in the low-mass regime) fit a conical shape to an outflow of ${\sim}450$~AU for a low-mass protostar, with an opening angle of $60\deg$.
If the outflow mechanism is indeed the same for low- and high-mass stars, and if this is a magnetic tower flow, then the outflow detected by \cite{pety_plateau_2006} should re-collimate at larger radii and later times, if accretion continues.

\cite{wu_study_2004} and \cite{beuther_massive_2002} find average opening angles of ${\sim}53 \deg$ over the same samples of $>10^3 \Lsol$ sources (corresponding to $5-15\Msol$ protostars in \citealt{wu_study_2004}) mentioned above, which are higher limits though, due to angular resolution and projection effects \citep{beuther_massive_2002}.
These are typically larger than what we obtain in runs \textsc{NoTurb} and \textsc{SubA}.
Therefore, this discrepancy may indicate a different outflow launching process, a smaller pressure confinement by the outer medium, or a need for higher numerical resolution at the outflow-environment interface in our simulation, if these values were to be confirmed with higher angular resolution studies.

\subsection{Alignment with magnetic fields, core-scale angular momentum and disk}
\label{sec:alignoutflow}

Let us now compare the values obtained in Sect.~\ref{sec:alignoutflow1} to observational studies, in both the low- and high-mass regimes, because, as we will see, so far there is no hint for a different orientation mechanism depending on the stellar mass.
In the low-mass regime, \cite{hull_misalignment_2013} observe that the angle distribution between outflows and magnetic fields on scales of ${\sim}1000$~AU is consistent with random distribution or preferentially perpendicular, on a sample of 16 low-mass protostars.
On the core-scale, \cite{hull_tadpol_2014} reached similar conclusion.
With a sample of four low-mass isolated protostars, 
\cite{chapman_alignment_2013} came to the opposite conclusion, with a positive correlation between the outflow axis and the magnetic fields direction.
Interestingly, \cite{galametz_sma_2018} show that the best alignment between the magnetic fields and the outflow axis is observed for sources with no large ($>100$~AU) disk nor multiplicity.
Finally, in the high-mass regime, \cite{arce-tord_outflows_2020} reach the same conclusions as \cite{hull_tadpol_2014}: their distribution is best fitted by either a $50-70\deg$ preferential orientation or a random orientation between the outflow and the magnetic fields.

To begin with, our results seem to favour a random inclination on small scales, as observed by \cite{hull_misalignment_2013}, since the outflow orientation is initially set by the disk orientation, which depends on the initial momentum carried by turbulence.
Second, the sample of \cite{chapman_alignment_2013} is most likely comparable to our non-turbulent run \textsc{NoTurb}, since they only focus on protostars that are isolated (e.g. B335, \citealt{olofsson_new_2009}), while we show in Paper I that turbulence favors the formation of multiple stellar systems.
Therefore, the positive correlation between the outflow axis and the magnetic fields in \cite{chapman_alignment_2013} agrees with our results.
Moreover, the present study is consistent with the observations of \cite{galametz_sma_2018}.
In fact, we only observe large rotating structures and multiple systems for super-Alfv\'enic runs, for which the outflow-magnetic field misalignment is indeed larger than in the sub-Alfv\'enic runs.
Overall, our work would suggest that the preferential perpendicular orientation ($>45\deg$) or random orientation would be obtained for systems with the Alfv\'enic Mach number $\mathcal{M_\mathrm{A}}>1$, as a consequence of the outflow being perpendicular to the disk, whose orientation is set by the initial angular momentum.
On the contrary, it would suggest that a better alignment is obtained for $\mathcal{M_\mathrm{A}}<1$, because the field line geometry or the streamers (perpendicular to $\mathbf{B}$) re-orient the outflows towards the core-scale magnetic field axis.

Finally, let us take a look at the magnetic field strength within the outflow.
As the outflow grows, its mean magnetic field strength decreases.
We measure a mean field strength of $15\, \mathrm{\, mG}$ in run \textsc{NoTurb} at the time when the outflow reaches ${\sim}\, 2000$~AU, $5\, \mathrm{mG}$ when it reaches ${\sim}\, 5000$~AU.
Using the Chandrasekhar-Fermi method with ALMA/Very Large Array (VLA) observations, \cite{hirota_magnetic_2020} obtained a value of $30\, \mathrm{mG}$ at $100-200$~AU in the outflows of the high-mass protostar Orion Source I.
Computing the average in the outflow at a height between $150$~AU and $250$~AU, we have a field strength of ${\approx}60\, \mathrm{mG}$ in run \textsc{NoTurb}, ${\approx}50\, \mathrm{mG}$ in run \textsc{SupA} (measuring it at late times), ${\approx}50-60\, \mathrm{mG}$ in run \textsc{SubA} (depending on the lobe) and ${\approx}50 \,\mathrm{mG}$ in run \textsc{SupAS} (in the transient outflow).
These are consistent within a factor of $2$ with \cite{hirota_magnetic_2020}.

\section{Discussion}
\subsection{Comparison with previous works}

The main result of this paper, namely a magnetic origin for massive protostellar outflows, is consistent with the work of C21, who included a similar physics and initial conditions but a FLD method to treat both stellar radiation and dust emission.
While the radiative force is underestimated with the FLD (by 2 orders of magnitude typically, \citealt{owen_radiative_2014}, \citealt{mignon-risse_new_2020}), they observe roughly three orders of magnitude between the Lorentz force and the radiative force.
Under the hypothesis that the radiative force does not interfere with the magnetic outflow launching (see Sect.~\ref{sec:outflowchannel}), their work demonstrated the magnetic origin of massive protostellar outflows up to ${\sim}20\Msol$.
With the present work, we show the validity of this hypothesis and confirm this result, with a larger participation from radiation, even at moderate masses ($5\Msol$, see Appendix~\ref{fig:app_fldhyout}).

We find that the presence of a turbulent velocity field delays and perturbs the launching of outflows, especially when the turbulence is super-Alfv\'enic.
This picture is consistent with the recent study of \cite{machida_failed_2020}, where ram pressure was caused by infalling gas at high accretion rates.
Our results indicate that mechanism remains the same as in the non-turbulent case, namely a magnetic outflow, but the radiative contribution is larger than in the non-turbulent case, partially because the outflow is delayed and launched at a larger stellar luminosity.
In the most turbulent case, a monopolar outflow forms, while the outflows are bipolar in all other runs.
This particular case shows the possibility of launching MHD outflows even when the orientation between the disk and the core-scale magnetic field is close to $90\deg$, in agreement with \cite{joos_influence_2013}.
This contrasts with \cite{ciardi_outflows_2010} who did not include turbulence but only misaligned rotation.
Hence, including initial rotation only may be an oversimplification regarding the processes affected by the angular momentum-magnetic field misalignment, since a realistic turbulent velocity field actually carries a non-regular distribution of angular momentum.
Consequently, the organization of magnetic field for launching outflows is delayed but is not prohibited.

Several clues point at a possible magneto-centrifugal jet in our simulations, such as the acceleration region coinciding with sub-Alfv\'enic velocities, and the criterion of \cite{seifried_magnetic_2012}.
As discussed in the high-resolution studies of \cite{banerjee_massive_2007} in the ideal MHD frame and \cite{kolligan_jets_2018} with non-ideal MHD, obtaining numerically converged results on the magneto-centrifugal mechanism requires sub-AU resolution (Sect.~\ref{sec:limit}).
However, the co-presence of a "slow" magnetic tower flow and "fast" centrifugal wind we obtain agrees with their work. 
Furthermore, the comparisons with observations we draw in Sect.~\ref{sec:outflowsobs} mainly arise from the two low-velocity components, the magnetic tower flow and the radiative outflow.
Deviation from the observed values could be attributed to the unresolved high-velocity jet for which further studies should be dedicated.

Let us compare the outflow mass and rate with numerical works. \cite{matsushita_massive_2017} with resistive MHD have explored several values for the ratio of the gravitational to magnetic energy (hence the accretion rate). After outflow launching, they obtain outflow masses nearly equal to the protostar's mass at all time.
Hence, for the typical protostar masses we obtain here, their outflow mass is typically $2-15\Msol$.
Nonetheless, they cover a timescale of only $10$~kyr, which is likely attributed to the Ohmic dissipation constraints.
Hence, they consider very high accretion rates, in order to reach a mass of a few tens of solar masses.
If we only consider their runs with an accretion rate of the order of a few $10^{-4} \Msolyr$, similar to ours, they obtain an outflow mass of ${\sim}4\Msol$ for a ${\sim}4\Msol$ central protostar, while the disk is becoming gravitationally-unstable and the outflow mass highly variable.
Nonetheless, as mentioned in Table~\ref{table:results}, we notice a delay of at least $8$~kyr (corresponding to at least $4\Msol$ accreted) between the sink formation and the outflow launching, which is not the case in \cite{matsushita_massive_2017} and may be related to different initial conditions.
Hence, while we should not directly compare their outflow mass with ours at a given time (or sink mass), the value of $4\Msol$ only gives an order of magnitude estimate, consistent with our work.
Finally, we compare our results to the ideal MHD study conducted by \cite{seifried_magnetic_2012}, which is one of the few works focusing on magnetic outflows in the massive star formation context.
They obtain mass outflow rates of $10^{-4}\Msolyr$ (and do not include turbulence), which agrees with our non-turbulent run \textsc{NoTurb}.

\subsection{Impact of ambipolar diffusion}

Let us first focus on the presence of outflows and whether ambipolar diffusion impacts it. In has been shown in C21 that magnetic outflows develop in the ideal MHD case (their run \textsc{MU5I}) and when ambipolar diffusion is included (their runs \textsc{MU5AD}, \textsc{MU2AD} and \textsc{MU5ADf}). Nevertheless, they show that the strong increase in magnetic pressure in the ideal MHD case kicks-off the primary sink particle, shutting-off the outflow launching during ${\sim}20$~kyr. In their study, this behavior is absent when ambipolar diffusion is accounted for. We confirm its absence here, with and without turbulence.

Second, let us investigate the magnetic field topology. In Paper I and in C21, the vertical component of the magnetic field has been found to dominate the inner regions of the disk when ambipolar diffusion is included, without turbulence. On the opposite, in the ideal MHD case, C21 find that the inner region is strongly dominated by the toroidal component of the magnetic fields. Nevertheless, the picture we obtain with ambipolar diffusion might change at sub-AU scales where the gas is ionized and the field weakened by diffusion processes, as found by \cite{vaytet_protostellar_2018} . This results in a generation of toroidal field by the disk differential rotation around the protostellar core. Overall, such aspects should be addressed with all non-ideal MHD effects (see e.g. \citealt{wurster_impact_2021}), and going down to second Larson core scales.

\subsection{Comparison with observations}

We have found agreement with CO observations regarding the outflow mass rate and momentum rate for cores of $100\Msol$.
A possibility is that our initial conditions, namely a massive core of $100\Msol$, are representative of the low-mass range of high-mass stars precursors.

Our results point to a correlation between the accretion plane and the outflow direction. Even though disk scales are not resolved in the recent study by \cite{goddi_multidirectional_2020}, they show how the outflow sudden change of orientation could reveal the accretion mode around massive protostars such as accretion streamers from multiple directions or a (small, $<100$~AU) disk plane changing with time (see also Paper I). Large-scale simulations and long-time integration are needed to address this question.

We obtain outflows with larger and smaller opening angles than observed jets and molecular outflows, respectively.
While this may open the possibility for other mechanisms than the one we explore, it could indicate that the outflow border requires higher resolution than offered here.
This could also be attributed to our outflow selection criteria, especially to our velocity threshold (in the vertical direction), required to avoid capturing isolated gas with a positive radial velocity but unrelated to outflows.
Nevertheless, a further lead would be to determine the role of ambient thermal pressure at collimating the flow, to see whether collimation depends on the initial ambient temperature ($20$~K) and to investigate how the outflow-environment interface depends on numerical refinement, but this is beyond the scope of the present work. 

\subsection{Limitations}
\label{sec:limit}

Our method contains several limitations.
First, we have used a hybrid scheme to treat separately the stellar irradiation from the ambient radiation, but with gray (i.e. frequency-averaged) methods for each component.
As discussed in \cite{kuiper_fast_2010}, such a gray treatment would under- or overestimate the effect of radiative pressure depending on the stellar spectrum, compared to a frequency-dependent (multigroup) scheme.
Nonetheless, this is a second-order effect, while we have determined regions where radiative acceleration and Lorentz acceleration differ by more than one order of magnitude (Fig.~\ref{fig:forcesratio}).
Therefore, our conclusions should not be affected by the gray approximation.

We have also considered idealized conditions for protostar formation with an isolated pre-stellar core, while several models have emerged to show that most massive stars may in a highly dynamical environment (see e.g., \citealp{vazquez-semadeni_high-_2009}, \citealt{peretto_global_2013}).
While these may not change our qualitative results, our study of the accretion rate and the outflow observables properties (mass rate, momentum rate, opening angles) should be extended in the frame of large-scale simulations.

Finally, we do not have the resolution to capture high-velocity (${\sim}\,300 \,\mathrm{km \,s^{-1}}$) MHD jets launched in the vicinity of the star with convergence. 
They may be necessary though, to reproduce the well-collimated outflows we have mentioned (see e.g., \citealt{moscadelli_water_2005}), while radiative force could contribute to their partial de-collimation.
Actually, they may entrain the ambient gas and fit the outflow momentum rate observed in CO \citep{arce_molecular_2007}.
In that respect, the development of a subgrid model for such jets is a first step (\citealt{kuiper_protostellar_2015}, \citealt{rosen_role_2020}).
Similarly, the inclusion of photoionization \cite{kuiper_first_2018} and longer-time integration (to reacher higher stellar masses, at which photoionization may dominate) are required.
We leave this to further work.

\section{Conclusions}

We have used four radiation-magnetohydrodynamical simulations with ambipolar diffusion and hybrid radiative transfer.
This allows us to avoid, on the one hand, the magnetic field strength overestimation of the ideal MHD framework, and on the other hand, the radiative force underestimation of the flux-limited diffusion method, to characterize the protostellar outflows in an unbiased way.
We have investigated the impact of turbulence and magnetic field strength on the outflow mechanism by considering a turbulent initial velocity field, varying the initial Mach number and Alfv\'enic Mach number.
Our results can be summarized as follows:
\begin{enumerate}
\item Outflows developed in all runs, but are delayed with super-Alfv\'enic turbulence, in comparison to sub-Alfv\'enic turbulence. They are mainly bipolar, but in the super-Alfv\'enic, supersonic turbulence run we only observe a transient, monopolar outflow such as the few observed ($16\%$ in \citealt{wu_study_2004}). To a larger extent, this brings to the forefront the importance of the environmental ram pressure in the outflow physics.
\item All outflows emerge from a magnetic mechanism. We find a magnetic tower flow \citep{lynden-bell_magnetic_1996} acceleration on the largest volumes while the radiative acceleration contribution is dominant close to the star. There are hints of a magneto-centrifugal acceleration near the outflow axis but this requires dedicated studies at higher-resolution to be confirmed.
\item The radiative force does not disrupt the field topology, at least up to ${\sim}\,10^5 \,\Lsol$ (${\sim}\,23\, \Msol$) in run \textsc{LRNoTurb}. 
\item In comparison with CO observations of massive star-forming regions, we find an overall agreement on the outflow mass rate and momentum rates for a similar core mass ($100\Msol$).  
\item We do not find clear agreement with observational constraints about opening angles in sub-Alfv\'enic turbulence runs. We produce outflows that are wider than the observed collimated jets, but more collimated than the wide-angle outflows observed although these are limited by observational resolution. In run \textsc{SupA}, the stellar motions in its binary system cause the outflows to widen. 
\item We do not find preferential outflow-magnetic fields alignment, except at large distances in the sub-Alfv\'enic run \textsc{SubA}. 
Outflows are first launched nearly perpendicular to the disk plane, and align within less than $40\deg$ with the core-scale angular momentum. These results predict a random outflow-magnetic fields misalignment if $\mathcal{M_\mathrm{A}}>1$ and a slightly better alignment for $\mathcal{M_\mathrm{A}}<1$.
\end{enumerate}

To sum up, these results show that the magnetic outflows are good candidates regarding the outflow mass, mass ejection rate and mass momentum rate measured in massive protostellar outflows.
On the contrary, they also show that the effect of the ambient gas of the outflow collimation is poorly known, and that, in the present study, magnetic outflows cannot reproduce the opening angles obtained from observations.
Although the radiative acceleration dominates close to the star, it seems insufficient to perturb the magnetic field topology enough to prevent MHD outflows from being launched.
Therefore, the only candidate (so far) to disrupt the field geometry is photoionization, as pointed out by \cite{peters_interplay_2011}, and should occur at later times than those considered here. 
We finally show that outflows preferentially develop perpendicular to the disk, but their orientation is highly-dependent on the ambient gas ram pressure.
\\

\begin{acknowledgements}
      This work was supported by the CNRS "Programme National de Physique Stellaire" (\emph{PNPS}). The numerical simulations we have presented in this paper were produced on the \emph{CEA} machine \emph{Alfvén} and using HPC resources from GENCI-CINES (Grant A0080407247). The visualisation of \ramses{} data has been done with the \href{https://github.com/nvaytet/osyris}{OSYRIS} python package.
\end{acknowledgements}

\bibliographystyle{aa} 
\bibliography{Zotero} 

\begin{appendix}
\section{Luminosity injection in the sink particle volume: outflows}
\label{sec:app_fldhy}
 
In this appendix, we investigate the influence of the radiative transfer method and of the kernel function to deposit the luminosity within the sink volume on the outflows.
This is motivated by the fact that part of the sink sits onto the inner disk region, hence a portion of the stellar radiation (which depends on the disk density, the resolution, the luminosity and the opacities) is locally absorbed before it has time to escape the sink volume.
This is a limitation of the hybrid approach, because the re-emitted radiation is treated with the FLD method instead of the M1.
Since the FLD method does not model properly stellar radiation in such anisotropic geometries and underestimates the radiative force, if one is interested in the temperature or dynamics of the outflows, one may want to circumvent this limitation so that stellar radiation can effectively escape from the sink volume with the M1 module.

\begin{figure*}
\centering
    \includegraphics[width=14cm]{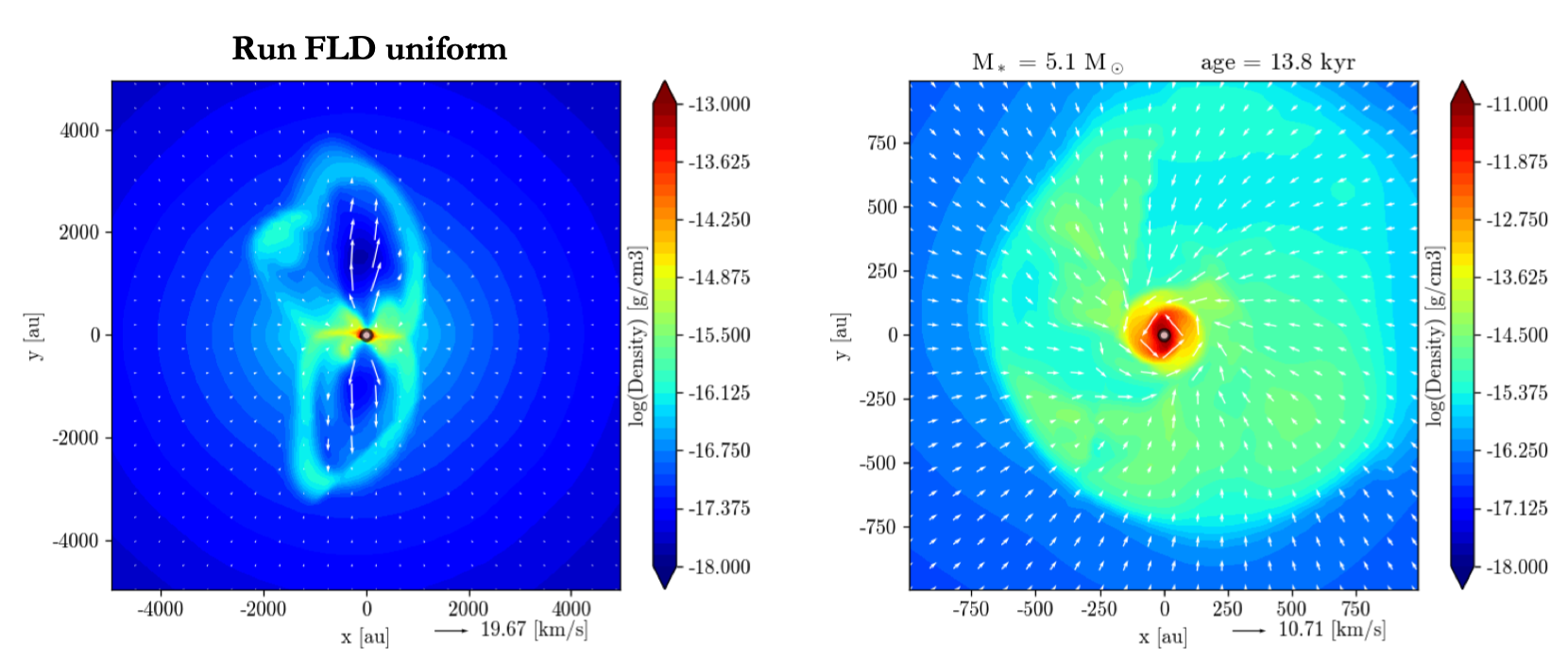}
    \includegraphics[width=14cm]{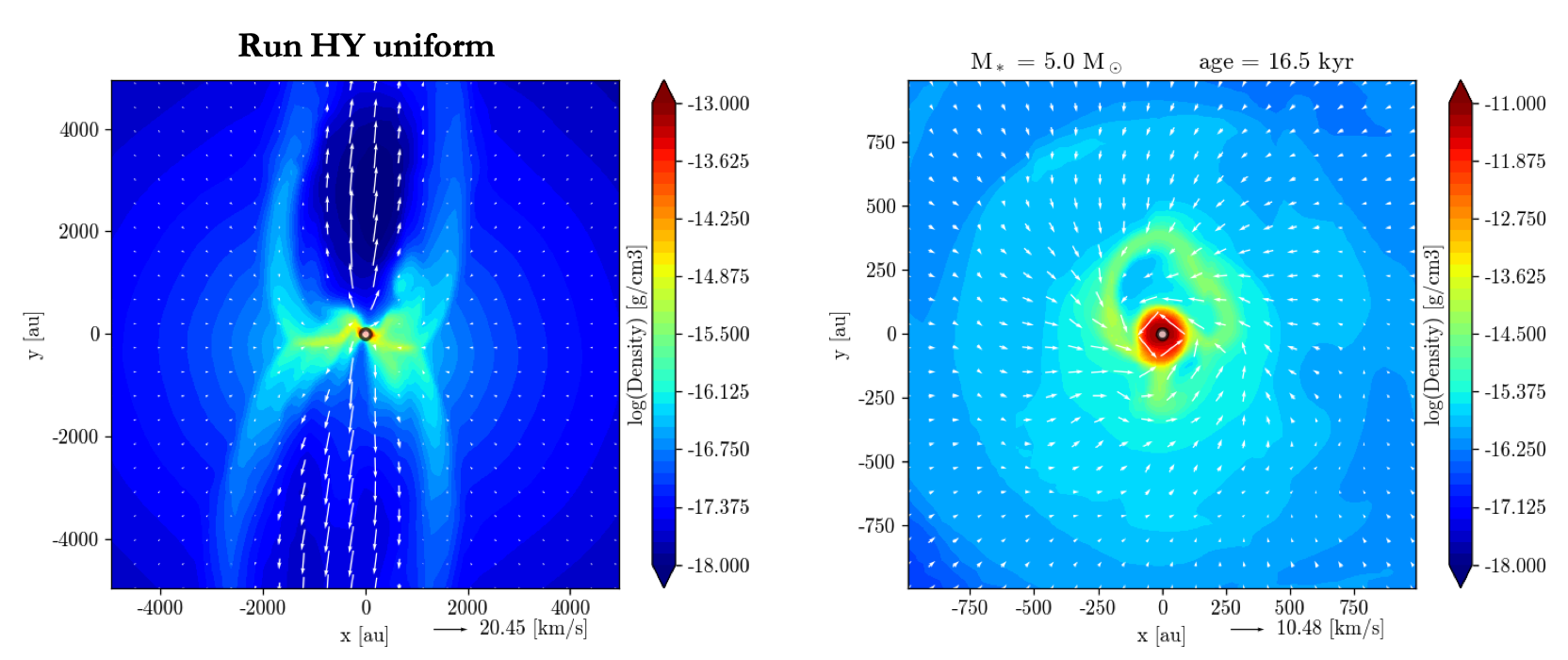}
    \includegraphics[width=14cm]{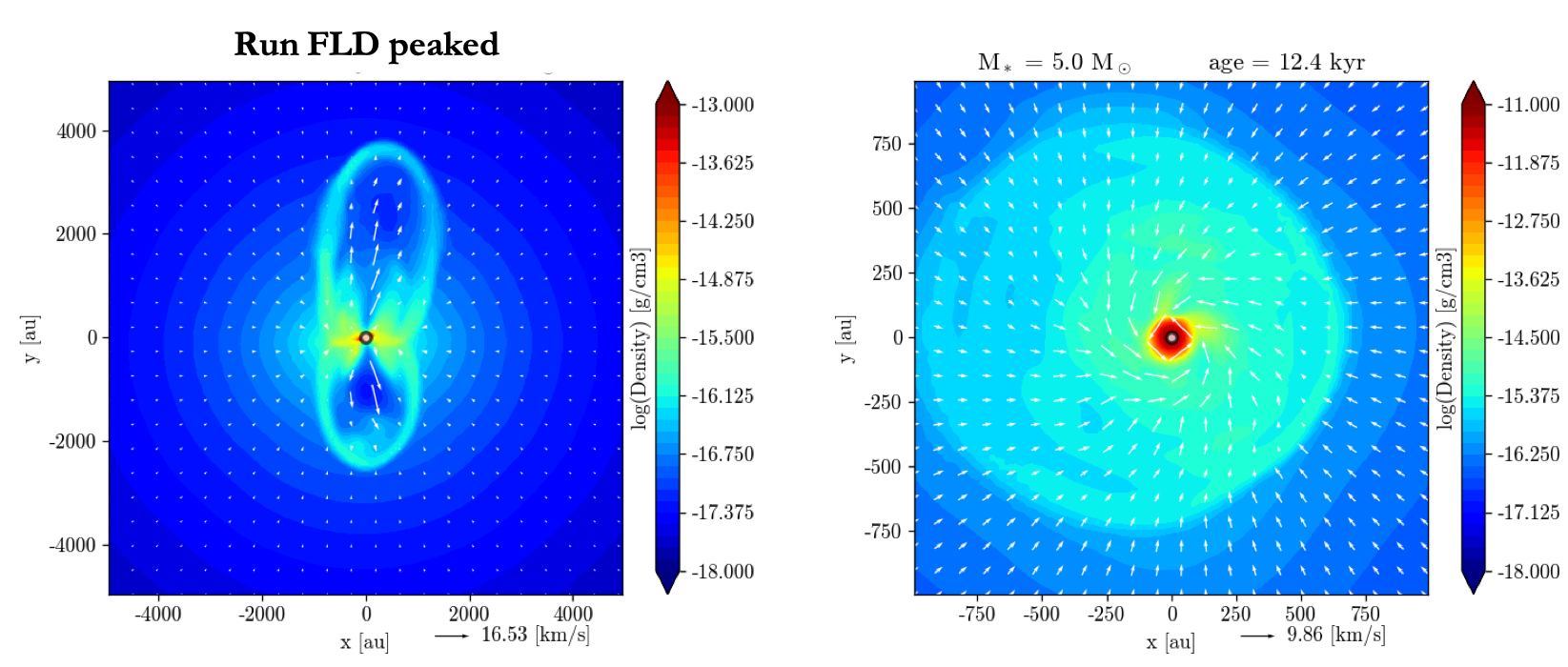}
    \includegraphics[width=14cm]{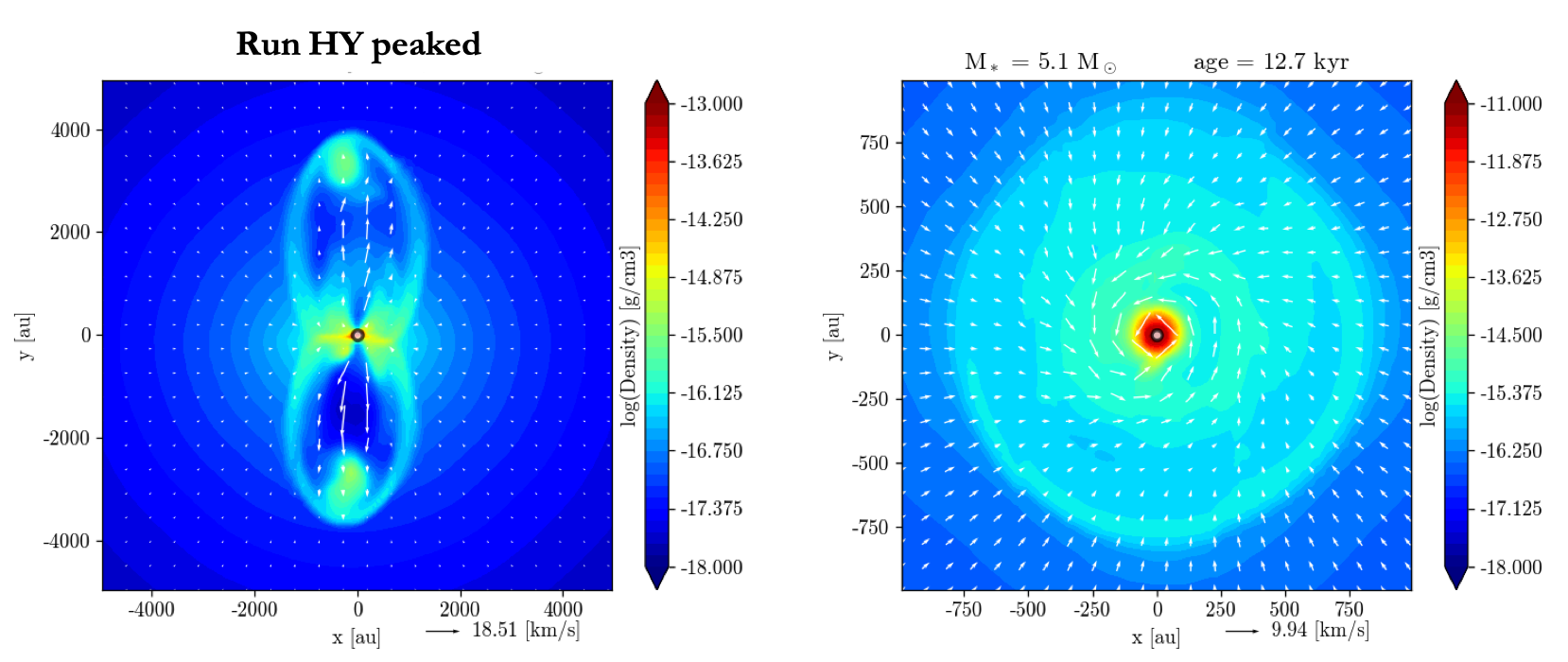}
    \caption{Density slices perpendicular (left) and parallel to the disk plane (right). First row: FLD method, uniform; second row: hybrid method, uniform; third row: FLD method, peaked; fourth row: hybrid method, peaked.}
    \label{fig:app_fldhyout}
\end{figure*}

The simulations are the same as run \textsc{NoTurb}: they include non-ideal MHD (ambipolar diffusion) but no turbulence.
Four simulations are considered: two with the flux-limited diffusion ("FLD") and two with the hybrid radiative transfer approach ("HY").
For each radiative transfer method, we test two injection kernels: either the luminosity is deposited uniformly over the sink volume ("uniform"), or only over the central oct ("peaked").

Figure~\ref{fig:app_fldhyout} shows the density slices perpendicular to the disk and in the disk plane, for each run, when the central star mass is ${\approx}5\Msol$.
The outflows are larger with the hybrid method than with the FLD, as expected from the radiative force estimations in \cite{mignon-risse_new_2020}.
They also appear less symmetric (with respect to the disk plane) in the FLD runs.
We note the presence of high-density "clumps" at the outflow front in the HY runs.
These are likely due to the greater acceleration by the stellar radiative force, compared to the FLD runs, which shocks with the outer medium.

Let us estimate the influence of the luminosity injection function.
For both radiative transfer methods, the "peaked" run leads to smaller outflows than the "uniform" run.
The difference in outflow size is even more obvious for the HY runs, because, as mentioned above, the M1 radiative force is significantly larger than the FLD radiative force.
Indeed, when all the luminosity is injected in the central oct, part of the radiation is absorbed and re-emitted with the FLD method, thus the outflow and disk properties can resemble that of the FLD runs.
On the opposite, the sink volume is larger than the local disk scale height, hence among the cells where luminosity has been injected uniformly there are some cells located outside the disk, so that stellar radiation can directly escape without being absorbed.
In that regard, a subgrid model with uniform injection reproduces one of the key features we are interested in.
Moreover, with such an injection method we find that the highest gas velocity is roughly $25\%$ smaller in run FLD run ($20\, \mathrm{km \, s^{-1}}$) than in the HY run ($26\, \mathrm{km \, s^{-1}}$), at ${\approx}5\Msol$, indicating that radiative acceleration is not negligible in the outflow cavity opened by magnetic processes.

Nevertheless, a uniform injection of luminosity within the sink volume is not physically satisfying.
In fact, the M1 radiative flux which powers the radiative force indirectly depends on the local radiative energy gradient.
If the injection is uniform over the sink volume, radiative energy is more absorbed in the central cells (which sit onto dense gas) than above and below the disk plane (where lower-density gas is located).
This results in a radiative flux oriented towards the central cells and consequently in a spurious radiative force oriented towards the central cells, from above and below the disk plane.
For this reason, we do not adopt a uniform luminosity injection function in this paper but rather set the sink volume as entirely optically-thin.
\end{appendix}

\end{document}